\documentclass[11pt,english]{article}
\usepackage{graphicx}
\usepackage{amstext}
\usepackage{amsbsy}
\usepackage{caption}
\usepackage{etoolbox}
\usepackage{makeidx}
\include{epsf}
\usepackage{amsfonts} 
\usepackage{authblk} 
\usepackage{sectsty}
\usepackage{amsmath,amssymb,epsfig}
\usepackage{amscd}
\usepackage{amsthm}
\usepackage{mathrsfs}
\usepackage{dsfont}
\usepackage[applemac]{inputenc}
\usepackage[english]{babel}
\usepackage{enumitem} 
\usepackage[]{latexsym}
\usepackage{braket}
\usepackage{caption}
\usepackage{hyperref}
\hypersetup{
pdftitle={},%
pdfauthor={},%
pdfsubject={},%
pdfkeywords={},%
colorlinks=true,%
linkcolor=blue,%
citecolor=red,%
linktocpage=true,%
hyperfootnotes=true,%
pageanchor=true
}

\newcommand{\be}{\begin{equation}}
\newcommand{\ee}{\end{equation}}
\def\beqa{\begin{eqnarray}}
\def\eeqa{\end{eqnarray}}
\def\bean{\begin{eqnarray*}}
\def\eean{\end{eqnarray*}}
\def\nn{ \nonumber}

\newcommand{\R}{\mathbb{R}}
\newcommand{\C}{\mathbb{C}}

\newcommand{\tg}{{\tilde{g}}}
\newcommand{\te}{{\tilde{e}}}
\newcommand{\tQ}{{\tilde{Q}}}
\newcommand{\tI}{{\tilde{I}}}
\newcommand{\tJ}{{\tilde{J}}}
\newcommand{\tX}{{\tilde{X}}}
\newcommand{\tY}{{\tilde{Y}}}
\newcommand\gl{\mathfrak}

\newcommand{\ad}{\hbox{ad}}

\newcommand{\eqn}[1]{(\ref{#1})}
\newcommand{\del}{\partial}
\newcommand{\Tr}[1]{\:{\rm Tr}\,#1}
\renewenvironment{thebibliography}[1]
         {\section*{References}\frenchspacing\small
          \begin{list}{[\arabic{enumi}]}
         {\usecounter{enumi}\parsep=2pt\topsep 0pt
         \settowidth{\labelwidth}{[#1]}
         \leftmargin=\labelwidth\advance\leftmargin\labelsep
         \rightmargin=0pt\itemsep=1pt\sloppy}}{\end{list}}

 \numberwithin{equation}{section}

\title{\textbf{Doubling, T-Duality and Generalized Geometry: \\ a Simple Model}
\\
\vspace{0.5cm}}
\date{}

\author[1,3]{Vincenzo E. Marotta}
\author[2]{Franco Pezzella}
\author[1,2]{Patrizia Vitale}

\affil[ ]{}
\affil[1]{\textit{\footnotesize Dipartimento di Fisica ``E. Pancini'', Universit\`a di Napoli Federico II, Complesso Universitario di Monte S. Angelo Edificio 6, via Cintia, 80126 Napoli, Italy.}}
\affil[2]{\textit{\footnotesize INFN-Sezione di Napoli, Complesso Universitario di Monte S. Angelo Edificio 6, via Cintia, 80126 Napoli, Italy.}}
\affil[3]{\textit{\footnotesize Department of Mathematics, 
Heriot-Watt University
Colin Maclaurin Building, Riccarton, Edinburgh EH14 4AS, U.K.}}
\affil[ ]{}
\affil[ ]{\footnotesize e-mail: \texttt{vm34@hw.ac.uk, franco.pezzella@na.infn.it, patrizia.vitale@na.infn.it}}
\begin{document}
\maketitle
\begin{abstract}
\small
A simple mechanical system, the three-dimensional isotropic rigid rotator, is here investigated as a 0+1 field theory, aiming at further investigating the relation between Generalized/Double Geometry on the one hand and Doubled World-Sheet Formalism/Double Field Theory, on the other hand.
The model is defined over the group manifold of $SU(2)$ and a dual model is introduced having  the Poisson-Lie dual of $SU(2)$ as configuration space. A generalized action %functional
with configuration space  $SL(2,\C)$, i.e. the Drinfel'd double of the group $SU(2)$, is then defined: it reduces to the original action of the rotator or to its dual, once constraints are implemented. The new action contains twice as many variables as the original. %as is also seen in Double Field Theory. 
Moreover its geometric structures can be understood in terms of Generalized Geometry.\\
{\it keywords: Generalized Geometry, Double Field Theory, T-Duality, Poisson-Lie symmetry}
\end{abstract}

\newpage
\tableofcontents

\section{Introduction}
Generalized Geometry (GG) was first introduced by N. J. Hitchin in ref. \cite{hitchin1}. As the author himself states in his pedagogical lectures \cite{hitchin2},  it is based on two premises: the first consists in replacing the tangent bundle $T$ of a manifold $M$ with  $T\oplus T^*$, a bundle with the same base space $M$ but  fibers given by the direct sum of tangent and cotangent spaces. The second consists in replacing the Lie bracket on the sections of $T$, which are vector fields, with the Courant bracket which involves vector fields and one-forms. The construction is then extended to general vector bundles $E$ over $M$ so to have $E\oplus E^*$ and a suitable bracket for the sections of the new bundle.  

The formal setting of GG has recently attracted the interest of theoretical physicists in relation to  Double Field Theory (DFT)  \cite{HZ}. We shall propose in this paper a model whose analysis can help to establish more rigorously a possible bridge between the two through the doubled world-sheet formalism that generates DFT.  

%DFT   aims at understanding  the behavior \textcolor{magenta}{under T-duality transformations of physical models, including strings,  described by field theories exhibiting such transformations as a symmetry of the dynamics; the symmetry however is not manifest at the level of the action functional. 
DFT has emerged  as  a proposal  to incorporate T-duality \cite{porrati, alvarez},  a peculiar symmetry of a compactified string on a $d$-torus $T^{d}$ in a $(G,B)$-background, as a manifest symmetry of the string effective  field theory.  In order  to achieve this goal, the action of this field theory has to be generalized in such a way that the emerging  carrier  space of the dynamics  be  {\em doubled} with respect to the original.  What makes T-duality a distinctive symmetry of  strings is that these latter, as extended objects and differently from particles, can wrap non-contractible cycles.  Such a wrapping implies the presence of winding modes that have to be added to the ordinary momentum modes which take integer values along compact dimensions. T-duality  is an $O(d,d;Z)$  symmetry of the dynamics of a closed string under, roughly speaking, the exchange of winding and momentum modes and establishes, in this way, a connection between the physics of strings defined on different target spaces. 

DFT is supposed to be an $O(d,d;Z)$ manifest space-time {\em effective} field theory description coming from a manifestly T-dual invariant formulation of a string world-sheet, i.e. from a {\em doubled world-sheet} \footnote{Let us observe here that we retain the name {\em doubled world-sheet} since this has become of common use, but %as we are going to explain below with more details,
 actually it is the string target-space which is {\em doubled} and not the world-sheet.}. In fact, 
a formulation of the world-sheet action of the bosonic string, in which T-duality is manifest, was already initially proposed in  ref.s \cite{Tseytlin, Duff} and, later, in  \cite{Hull, Berman, park, Copland} (see also more recent works in  \cite{pezzella, Bandos, nibbelink, Ma}). This string action must contain information about windings and therefore it is based on two sets of coordinates: the usual ones $x^{a}(\sigma, \tau)$ and the ``dual'' coordinates $\tilde{x}_{a} (\sigma, \tau)$, $(a=1,...,d)$  conjugate to the winding modes. %Hence, to each compact space  coordinate characterized by a certain winding, one can associate  a new coordinate conjugate to the winding it is based .
  In this way the $O(d,d;Z)$ duality becomes  a manifest symmetry of the world-sheet action. 
A corresponding doubling of {\em all} the $D$ space-time degrees of freedom  (vielbeins in this case, not only relatively to the compact dimensions) in the low-energy effective action first occurred in ref. \cite{siegel}  where, a manifestly $O(D,D;R)$ form of the target-space effective action was obtained, and such symmetry was realized linearly, even at the price of loosing manifest Lorentz invariance (in target-space). In a sense, this can be considered as a pionering work on what would be later defined Double Field Theory, where the coordinates of the carrier space-time, that are nothing but that the {\em fields} on the string world-sheet, are doubled in order to have a T-duality symmetric field theory.

% Then, a T-duality symmetric field theory  A possible way to implement such information consists in introducing a new set of dual coordinates in target space \textcolor{blue}{ (that is {\it fields} on the string world sheet)}, canonically conjugate to the winding modes. .  It is then possible to describe the dynamics in the Hamiltonian setting by means of Poisson brackets which generalize the canonical ones to such an extended setting. In this sense DFT is a  {\em double} theory: it doubles the coordinates of the compact space.  But actually the doubling can involve the whole $D$-dimensional target space, since one can also formally associate  to each non-compact dimension its own dual coordinate, on which no physical quantities really depend. 

Despite the preamble, which gives credit to the strings related literature for focusing on the geometrical content of the doubled world-sheet and DFT, the interest  for the subject % is naturally not limited to string theory but  is 
is relevant in the broad area of field theory when one deals with duality symmetries of the dynamics which are not manifest at the level of the action.%For an early paper see for example  \cite{siegel84}, where  Siegel addresses the problem of constructing manifestly Lorentz invariant actions for self dual gauge fields by means of the introduction of auxiliary fields in the Lagrangian.

A few remarks that clarify the philosophy of the paper are here in order. 
First of all, it is worth stressing again that, in the framework of string theory,  the doubling takes place in the $D$-dimensional target space $M$ of the non-linear sigma model underlying the string action, by introducing  new fields $\tilde x_i (\sigma,\tau)$, which are dual to $x^i(\sigma,\tau)$, with $i = 1, \dots, D$. From this point of view, a first analogy with Generalized Geometry is straightforward, by identifying $x^i, \tilde x_i $ with sections of a generalized bundle $E\oplus E^*$ over the world sheet of the string. Secondly, it is only when the target space is considered as the configuration space of the {\it effective field theory} we are going to deal with, that the doubling is reinterpreted as a doubling of the configuration space. Actually,  the original non-linear sigma model has no doubled coordinates, but what is doubled are the field coordinates. 
 When the effective field theory derived from the Polyakov string action is considered, then the dual fields $x^i, \tilde x_i$ are seen as coordinates of the carrier space of the effective dynamics, which corresponds to the string target space. DFT is thus  formulated in terms of the background fields $G_{ij}$  (the target-space metric tensor) and $B_{ij}$ (the Kalb-Ramond field), with $i,j = 1, \dots, D$, in addition to a dilaton scalar field $\phi$. These fields depend, in that framework, on doubled coordinates $x^{i}$ and $\tilde{x}_{i}$ even if there is no doubling of their tensor indices. The gauge symmetry parameters for DFT are  the vector fields  $\xi^{i} (x, \tilde{x})$, which parametrize diffeomorphisms and are sections of  the tangent bundle of the doubled manifold, together with the one-forms $\tilde{\xi}_{i} (x, \tilde{x})$, which describe gauge transformations of the Kalb-Ramond field $B_{ij}$ and are sections of the cotangent bundle of the doubled manifold. When considering vector fields and one-forms as components of a generalized (indeed doubled) vector field on the carrier space of the effective dynamics (which is itself doubled), one has,  on the one hand,  another instance of field doubling, but on the other hand, at the same time, a  section of a generalized tangent bundle as in Generalized Geometry. The precise mathematical meaning of considering $\xi^{i} $ and $\tilde{\xi}_{i}$ on the same footing amounts to defining  generalized Lie brackets, which encode a  mutual non-trivial action of one onto the other  \cite{HZ2}. These are the so-called $C$-brackets, first introduced, together with other relevant aspects of DFT, in ref.s \cite{siegel}. 
 $C$-brackets provide an $O(D,D)$ covariant, DFT generalization of  Courant brackets. More precisely, it can be shown that they  reduce to Courant brackets if one drops the dependence of the doubled fields on the coordinates $\tilde{x}_{i}$. 
 The geometry of the effective dynamics is thus more appropriately renamed {\em Doubled Geometry} (DG).

To summarize, doubling can emerge at  different stages:
\begin{itemize} 
\item at the the level of fields on a given configuration space, for example   the sigma-model fields  $x^i,\tilde x_i$ both depending on the world sheet coordinates $(\sigma,\tau)$;
\item at the level of configuration space coordinates, with fields $\phi$ depending on twice the initial configuration space variables,  $\phi= \phi(x^i, \tilde x_i)$;
\item at the level of both, fields and coordinates: an example is provided by the gauge fields $\xi^i(x, \tilde x), \tilde\xi_i(x, \tilde x)$.  
\end{itemize}
There is therefore an interplay between GG and  DG on the one hand and doubled world-sheet and DFT on the other hand  which, within  the framework we have sketched, emerges from the identification of the appropriate carrier space of the dynamics. 
Such interplay does not involve only the above mentioned T-duality  to which one usually refers as Abelian T-duality,  but it could be enlarged also to the other two dualities connecting non-linear sigma models, the non-Abelian T-duality and the Poisson-Lie T-duality. The term ``Abelian T-duality" refers to the presence of global Abelian isometries in the target spaces of both the paired sigma-models \cite{buscher, rocek} while ``non-Abelian" refers to the existence of a global non-Abelian isometry on the target space of one of the two sigma-models  and of a global Abelian isometry on the other  \cite{quevedo}.  The ``Poisson-Lie T-duality" generalizes the previous definitions to all the other cases, including the one of a dual pair of sigma models both having non-Abelian isometries in their target spaces \cite{Klim, Klim2}. More easily, the classification of T-dualities is given by the types of underlying Drinfel'd doubles: Abelian doubles for the Abelian T-duality, semi-Abelian  doubles for the non-Abelian T-duality and non-Abelian doubles for the Poisson-Lie T-duality.

 It is then clear that models whose carrier space  is a Lie group $G$ can be very helpful in better understanding the above mentioned relation in all these cases, because the notion of dual of a Lie group is well established together with that of double Lie group  and the so called Poisson-Lie symmetries \cite{drinfel'd, semenov}. The idea of investigating such geometric structures in relation to duality in field theory has already been applied to  sigma models by Klim\v{c}\'ik and \v{S}evera in \cite{Klim} (also see \cite{sfetsos}, \cite{falceto}) where the authors first introduced the notion of Poisson-Lie T-duality. Since then, there has been an increasing number of papers in the literature, focusing on Poisson-Lie dual sigma models (see for example ref. \cite{lledo}).
 On the other hand,  in  ref. \cite{AF}, the phase space $T^*G$ was already proposed as a toy model for discussing conformal symmetries of chiral models, in a mathematical framework which is very similar to the one adopted here. Double Field Theory on group manifolds, including its relation with Poisson-Lie symmetries, has been analyzed in \cite{hassler}. In the present paper,  we propose a fresh look at the subject in relation to the recent developments in GG and DFT by studying a model, the three-dimensional isotropic rigid rotator (IRR) that provides a one-dimensional simplification of a sigma model which can be doubled in order to have a manifestly Poisson-Lie duality invariant doubled world-sheet. 
  
 This is the first of a series of two papers.  We study the IRR having as configuration space the group manifold of $SU(2)$ and introduce a model on the dual group $SB(2,\C)$. Their properties under Poisson-Lie transformations are considered  as an extended model on the double group, the so-called classical Drinfel'd double $SL(2,\C)$, that is formulated in terms of a generalized action, which we shall refer to as the parent action. In particular, we emphasize how a natural para-hermitian structure emerges on the Drinfel'd double and can be used to provide a ``doubled formalism'' for the pair of theories.
An alternative description of the IRR model on the Drinfel'd double was already proposed in \cite{marmo:articolo1}, although no  dual model was introduced there, being the accent on the possibility of describing the same dynamics with a different phase space, the group manifold $SL(2,\C)$, which relies on the fact that the latter is symplectomorphic to the cotangent bundle of $SU(2)$ \cite{MI98}.  
 
Since our model describes an example  of  particle dynamics, the most appropriate doubling within those enumerated above is  the doubling of the configuration space. For the same reason, we shall see that 
  the model considered here is too simple to exhibit symmetry under duality transformation, although a generalization to field theory is possible. Indeed,  we may look at  the model   as a $0+1$ field theory, thus paving the way for a genuine $1+1$ field theory, the $SU(2)$ principal chiral model, which, while being modeled on the IRR system, will exhibit interesting properties under duality transformations. This will be briefly discussed  in the concluding section while the model will be analyzed in detail in a forthcoming paper \cite{MPV2}. 

 The paper is organized as follows. In Sect. \ref{rigidrot}  the dynamics of the IRR on the group manifold of the group $SU(2)$ is reviewed. In Sect.  \ref{drinfel'd}  an account of the mathematical framework that is going to be used is given, with Poisson-Lie groups and their Drinfel'd doubles discussed in some detail. In  Sect. \ref{dualrot} a model on the dual group of $SU(2)$, the group $SB(2,\C)$,  is introduced  and its dynamics analyzed. The two models are seen to be dual to each other in a precise  mathematical meaning:  their configuration spaces are  dual partners in the description  of the group $SL(2,\C)$ as a Drinfel'd double. Moreover, the role of the two groups can be exchanged in the construction of the double and each model exhibits a global symmetry with respect to the action of the dual group.
In Sect. \ref{gensec},  a dynamical model on the Drinfel'd double is proposed: it has doubled configuration variables with respect to the original IRR coordinates, and doubled generalized momenta $(I_i,\tI^i)$ whose  Poisson brackets can be related to Poisson-Lie brackets on the two dual groups. The full Poisson algebra of momenta is isomorphic to the algebra of $SL(2,\C)$, namely a semisimple group, with each set of momenta underlying  a non-Abelian algebra. That is why we refer to the two models as non-Abelian duals giving rise, according to the above mentioned definitions, borrowed from the existing literature, to a Poisson-Lie T-duality. 
 %Moreover, the latter  can be interpreted as tangent and cotangent space coordinates on the generalized $T\oplus T^*$ bundle over $SU(2)$ (or, dually, over the group $SB(2,\C)$), with Poisson brackets yielding a Poisson realization of C-brackets. 
In Sections \ref{standardlag}, \ref{recdu}, we address the problem of recovering the IRR model and its dual. The generalized, or parent action, exhibits global symmetries, which can be gauged, as is customary in DFT (see for example \cite{Hull, park}). It is  proven that, once chosen a parametrization for the group $SL(2,\C)$, by gauging the left $SB(2,\C)$ symmetry the IRR model is retrieved, whereas, by gauging the right $SU(2)$  symmetry the dual model is obtained.
In Sect. \ref{hamform}, we introduce the Hamiltonian formalism for the double model and in \ref{canform} we study in detail the full Poisson algebra, together with  the  Hamiltonian vector fields associated with momenta $(I_i,\tI^i)$. The latter  yield  an algebra which is closed under Lie brackets, which can be seen as {\it derived} C-brackets \cite{deser1, deser2}. In Sect. \ref{PLsym} we discuss in some detail to what extent the two models introduced exhibit Poisson-Lie symmetries. 
Finally, in Section \ref{concl} we outline the generalization to 1+1 dimensions for the principal chiral model and give our conclusions.

 While completing the article we have become aware of the work in refs. \cite{lust, KS18}. In the first one   non-Abelian T-duality is analyzed within the same mathematical framework, whereas the latter studies an interesting mechanical model, the electron-monopole system, within the DFT context. Their relation  with the present work should be further investigated; we plan to come back to this issue in the future.

 \section{The Isotropic Rigid Rotator}\label{rigidrot}
The Isotropic Rigid Rotator (IRR) provides   a classical example of dynamics on a Lie group, the group being in this case $SU(2)$ with its cotangent bundle $T^*SU(2)$, the carrier space of the Hamiltonian formalism, carrying  the group structure of a semi-direct product. In this section, the Lagrangian and Hamiltonian formulations of the model on the group manifold are reviewed.  Although being simple, the model captures relevant characteristics of   the dynamics of many interesting physical systems, both in particle dynamics  and in field theory, such as Keplerian systems, gravity in 2+1 dimensions in its first order formulation, with and without cosmological constant \cite{witten}, Palatini action with Holst term \cite{Holst}, and principal chiral models \cite{gursey}.

\subsection{The Lagrangian and Hamiltonian Formalisms}
As carrier space for the dynamics of the three dimensional rigid rotator in the Lagrangian [Hamiltonian] formulation  we can  choose  the tangent [cotangent] bundle of the group $SU(2)$. We follow  ref. \cite{marmosaletan} for the formulation of the dynamics over Lie groups.

A suitable action for the system  is the following
\be
S_0= \int_\R L _0~dt=-\frac{1}{4} \int_\R \Tr ( g^{-1} d g\wedge * g^{-1} dg) =-\frac{1}{4}\int_\R  \Tr (g^{-1}{\dot g})^2  dt \label{lag}
\ee
with $g:t\in \R\rightarrow SU(2)$, the group-valued target space coordinates, so that 
\be
g^{-1} d g = i \alpha^k \sigma_k  \nonumber
\ee
 is the Maurer-Cartan left-invariant  one-form, which is Lie algebra-valued, $\sigma_k $ are the Pauli matrices, $\alpha^k$ are the basic left-invariant one-forms, $*$ denotes the Hodge  
star operator on the source space $\R$, such that $* dt = 1$,  and $\Tr$ the trace over the Lie algebra. Moreover,  $g^{-1}{\dot g}$ is the contraction of the Maurer-Cartan one-form with the dynamical vector field $\Gamma=d/dt$,  $g^{-1}{\dot g}\equiv (g^{-1}{ d g}) (\Gamma)$.  Let us remind here that the Lagrangian is written in terms of the non-degenerate invariant scalar product defined on the $SU(2)$ manifold and given by $\langle a | b\rangle = \mbox{Tr}(ab)$ for any two group elements. The model can be regarded as a $(0+1)$-dimensional field theory which is group-valued. 

The group manifold can be parametrized with  $\R^4$ coordinates, so that  
$g\in SU(2)$ can be read  as $g= 2(y^0 e_0 
+i  y^i {e_i})$, with $(y^0)^2+ \sum_i (y^i)^2=1$, $e_0={\mathbb I}/2$, $e_i=  \sigma_i/2$ the $SU(2)$ generators.  One has then:
\be
y^0=  \Tr  (g e_0), \;\;\; y^i=-{i} \Tr (g e_i) \; \; \;  i=1, \dots ,3   \nonumber
\ee
By observing that 
\be
g^{-1}  \dot g  =  i (y^0 \dot y^i-y^i \dot y^0+  {\epsilon^{i}}_{jk} {y^j \dot y^k })\sigma_i   = i \dot{Q}^{i} \sigma_{i} \label{qdot}
\ee
we define the left generalized velocities $\dot Q^i$ as 
\be
\dot Q^i  \equiv
 (y^0\dot y^i-y^i \dot y^0+  {\epsilon^{i}}_{jk} y^j \dot y^k) . \label{genvel}
\ee
$(Q^i, \dot Q^i)$ $i=1, \dots ,3$ are therefore  tangent bundle coordinates, with $Q^i$ implicitly defined. Starting  with a Lagrangian written in terms of right-invariant one-forms,  one could define right generalized velocities in an analogous way. They  give an alternative  set of coordinates  over the tangent bundle. 

The Lagrangian $L_{0}$ in eq. \eqn{lag} can be rewritten as:
\be
L_0=\frac{1}{2} (y^0\dot y^i-y^i \dot y^0+  {\epsilon^{i}}_{kl} y^k \dot y^l)(y^0\dot y^j-y^j \dot y^0+  
{\epsilon^{j}}_{mn} y^m \dot y^n)\delta_{ij} =\frac{1}{2} \dot Q^i \dot Q^j \delta_{i j}.  \nonumber
\ee
In the intrinsic formulation,  which is especially relevant in the presence of non-invariant Lagrangians, the Euler-Lagrange equations of motion are represented by:
\be
 {\sf L}_\Gamma  \theta_{L} -d L_0=0   \nonumber
\ee
being
\be
\theta_L= \frac{1}{2}\Tr[ g^{-1} \dot g  \;g^{-1} d g]= \dot Q^i \alpha^j \delta_{i j }    \nonumber
\ee
the Lagrangian one-form and ${\sf L}_{\Gamma}$  the Lie derivative with respect to $\Gamma$. 
By projecting along the basic left-invariant vector fields $X_i$ dual to $\alpha^i$, one obtains:
\be
i_{X_i} [ {\sf L}_\Gamma  \theta_{L} -d L_0]=0    \nonumber
\ee
Since $ {\sf L}_{\Gamma}$  and $i_{X_i}$ commute over the Lagrangian one-form,  one gets:
\be
{\sf L}_\Gamma(\dot Q^j\, i_{X_i}\alpha^l )\delta_{jl} - {\sf L}_{X_i} L_0 = 0  \nonumber
\ee
which implies
\be
{\sf L}_\Gamma\dot Q^j \delta_{ji} - \dot Q^p \dot Q^q {\epsilon_{ip}}^k\delta_{qk} = {\sf L}_\Gamma\dot Q^j \delta_{ji}= 0 \label{eqmo}
\ee
because of the rotation invariance of the product and the antisymmetry of the structure constants of $SU(2)$ as a manifestation of the  invariance of the Lagrangian under rotation.

Equivalently, the equations of motion can be rewritten as:
\be
\frac{d}{dt} \left(g^{-1}\frac{dg}{dt}\right)=0\label{eom}
\ee
being, from   eq. \eqn{qdot}:
\be
\delta_{ij} \dot{Q}^j= -{i}\Tr( g^{-1} \dot g \; e_i) .
\ee
Cotangent bundle coordinates can be chosen to be $(Q^i, I_i)$ with the $I_i$'s denoting the left momenta: 
\be
I_i= \frac{\del  L_0}{\del \dot Q^i}= \delta_{i j} \dot Q^j   \nonumber
\ee
An alternative set of fiber coordinates is represented by the right momenta, which are defined in terms of the right generalized velocities. 

The Legendre transform from $TSU(2)$ to $T^*SU(2)$ yields  the Hamiltonian function:
\be
H_0=[I_i \dot Q^i -L_{0}]_{\dot Q^i=\delta^{ij} I_j }= \frac{1}{2} \delta^{ij}I_i I_j \label{h0}  \,\,.
\ee
By  introducing  a dual basis $\{{e^i}^*\}$ in the cotangent space, such that $\langle{e^i}^*|e_j\rangle =\delta^i_j$, one can consider the linear combination:
\be
I=i\; I_i {e^i}^*. \label{Iform}
\ee 
 The dynamics of the IRR is thus  obtained from the Hamiltonian \eqn{h0} and   the   following Poisson brackets
\beqa
\{y^i,y^j\}&=&0\label{pp}\\
\{I_i,I_j\}&=& {\epsilon_{ij\;}}^k I_k \label{xx}\\
\{y^i,I_j\}&=&\delta^{i}_j y^0 +{\epsilon^i}_{\; jk}y^k \;\;\;  {\rm or ~ equivalently}\;\;\; \{g, I_j\}= 2 i  g e_j \label{ij}
 \eeqa
 which are derived from the first-order formulation of the action functional 
 \be
 S_1= \int \langle I | g^{-1}\dot g\rangle dt - \int H_{0} \, dt \equiv \int \vartheta -\int H_{0} dt   \nonumber
 \ee
 with $\theta$ the canonical one-form. 
 Indeed the symplectic form $\omega$ is readily obtained as 
 \be
 \omega= d \vartheta=  d I_i \wedge \delta^{i}_j  \alpha ^j - \frac{1}{2}I_i \delta_j^i {\epsilon^j}_{kl} \alpha^k\wedge \alpha^l   \nonumber
 \ee
 with  $ d\alpha^k= \frac{i}{2}\,{\epsilon^{k}}_{ij}\alpha^i\wedge\alpha^j$.  By inverting $\omega$ one finds the Poisson algebra \eqn{pp}-\eqn{ij}.  

The fiber coordinates  $I_i$  are   associated with the angular momentum components and the base space coordinates $(y^0, y^i)$ to the orientation of the rotator. The resulting system is rotationally invariant since
$ 
\{I_i, H_0\} = 0. $ 

The Hamilton equations of motion for the system  are:
\be 
\dot I_i= 0,\;\;\; g^{-1}\dot g= 2 i I_i \delta^{ij} e_j.  \nonumber
\ee
Thus the angular momentum
$I_i$ is a  constant of motion, while $g$ undergoes a uniform
precession. Since the Lagrangian and the Hamiltonian are invariant under right and left $SU(2)$ action, as well-known right momenta are conserved as well, being the model super-integrable.

 Let us remark here that, while the fibers of the tangent bundle $TSU(2)$ can be identified, as a vector space,  with the Lie algebra of $SU(2)$, $\mathfrak{su}(2)\simeq \R^3$, with $\dot Q^i$ denoting vector fields components, the fibers of the cotangent bundle $T^*SU(2)$ are isomorphic to the dual Lie algebra $\mathfrak{su}(2)^*$. As a vector space this is again $\R^3$, but the $I_i$ 's are now components of one-forms. This remark is relevant in the next section, when the Abelian structure of $\mathfrak{su}(2)^*$ is deformed.
 
As a group, $T^*SU(2)$ is the semi-direct product of $SU(2)$ and the Abelian group $\mathbb{R}^3$, with the corresponding Lie algebra given by: 
\beqa
\left[L_i,L_j\right]  &=&i {\epsilon_{ij}}^k L_k \label{JJ}\\
\left[T_i,T_j\right] &=& 0 \label{PP}\\
\left[L_i,T_j\right] &=&i {\epsilon_{ij}}^k T_k. \label{JP}
\eeqa
Then, the  non-trivial Poisson bracket on the fibers of the bundle, \eqn{xx},  can be 
understood in terms of the coadjoint action of the group $SU(2)$ on its dual algebra
$\mathfrak{su}(2)^*\simeq \mathbb{R}^3$  and it reflects the
non-triviality of the Lie bracket \eqn{JJ}. In this picture the Lie algebra generators $L_i$'s are
identified with the linear functions on the dual algebra.\footnote{The group semi-direct product structure of  the phase space $T^*SU(2)$ has  been widely investigated in literature in many different contexts. Besides classical, well known applications, some of which we have already mentioned in the introduction, let us mention  here applications in noncommutative geometry in relation  to the quantization of the hydrogen atom \cite{duflo}, to the electron-monopole system \cite{noi}, and, recently, to models on  three-dimensional space-time with $\mathfrak{su}(2)$ type non-commutativity \cite{vitalewallet, kupr, wallet, zoupanos}}.  

Before concluding this short review of the canonical formulation of the dynamics of the rigid rotator, let us stress the main  points which we are going to elaborate further:
\begin{itemize}
\item The carrier space of the Hamiltonian dynamics is represented by  the semi-direct product of a non-Abelian Lie group, $SU(2)$, and the Abelian group $\R^3$ which is nothing but the dual of its Lie algebra.
\item The Poisson brackets governing the dynamics are the Kirillov-Souriau-Konstant  brackets induced by the coadjoint action.
\end{itemize}
It has been shown in ref. \cite{marmo:articolo1} that the carrier space of the dynamics of the rigid rotator can be generalized to the semisimple group $SL(2, C)$, which is obtained by replacing the Abelian subgroup $R^3$ of the semi-direct product above, with a non-Abelian group.  The generalization is obtained by considering the {\it double} Lie group of $SU(2)$.  In this paper  such generalization will be further pursued giving rise to a giving rise to  the simplest instance of doubled dynamical model, together with its double geometry. The underlying mathematical construction of  Drinfel'd double Lie groups and their relation with the structures of Generalized Geometry is the subject of the next section.

\section{Poisson-Lie Groups and the Double Lie Algebra $\mathfrak{sl}(2,\mathbb{C})$}\label{drinfel'd}
In this section  we shortly review the mathematical setting of Poisson-Lie  groups and Drinfel'd doubles, see~\cite{semenov,alex,wein} for details, with the aim of introducing, in the forthcoming sections, new Lagrangian and Hamiltonian formulations of the IRR with a  manifest symmetry under duality transformation.  More precisely, in Section \ref{dualrot},  a model which is dual to the one described in Section \ref{rigidrot}  is introduced, while in Section \ref{gensec}  \textcolor{magenta}{ a new model is built}  with doubled dynamical variables  and with a manifest symmetry under duality transformation. The dynamics derived from the new action describes two models, dual to each other, being one the ordinary rigid rotator, the other a``rotator-like" system, with the rotation group $SU(2)$ replaced by its Poisson-Lie dual, the group $SB(2,\C)$ of Borel $2\times 2$ complex matrices. 
  
 { A
Poisson-Lie group is a Lie group \textcolor{magenta}{$G$ }equipped with a Poisson structure
which makes  the product $\mu :G\times G \rightarrow G$ a Poisson map if
$G \times G$ is equipped  with the product Poisson
structure. Linearization of the Poisson structure at the unit $e$ of
$G$ provides   a Lie algebra structure over the dual algebra  ${\gl g}^*=T^*_{e}(G)$ by the
relation
\begin{equation} 
\label{Liedual}
[d\xi_{1}(e),d\xi_{2}(e)]^{*}=d\{\xi_{1},\xi_{2}\}(e)
\end{equation}
with $\xi_i\in C^\infty(G)$. 
The compatibility condition between the Poisson and Lie structures  of $G$ yields the  relation:
\begin{equation} 
\label{comp} 
\left< [X,Y],[v,w]^*\right>+\left<\ad_v^*X,\ad_Y^*w\right>
-\left<\ad_w^*X,\ad_Y^*v\right>
-\left<\ad_v^*Y,\ad_X^*w\right>+\left<\ad_w^*Y,\ad_X^*v\right>=0\,
\end{equation} 
with $v,w\in\mathfrak{g}^*, X,Y\in \mathfrak{g}$ and $ \ad_X^*, \ad_v^*$ the coadjoint actions of the Lie algebras $\mathfrak{g}, \mathfrak{g}^*$ on each other.
This allows one to define a Lie bracket in ${\gl g}\oplus {\gl g}^*$ through the formula:
\begin{equation} 
\label{Liesuma}
[X+\xi,Y+\zeta]=[X,Y]+[\xi,\zeta]^{*}-ad^{*}_{X}\zeta + ad^{*}_{Y}\xi
+ ad^{*}_{\zeta}X - ad^{*}_{\xi}Y \,\, .
\end{equation}
If $G$ is connected and simply connected,~(\ref{comp}) is enough to
integrate $[\ ,\ ]^*$ to a Poisson structure on $G$ that makes it
Poisson-Lie and the Poisson structure is unique. The symmetry between
${\gl g}$ and ${\gl g}^*$ in~(\ref{comp}) implies that one has also a
Poisson-Lie group $G^*$ with Lie algebra $({\gl g}^*,[\ ,\ ]^*)$ and a
Poisson structure whose linearization at $e\in G^*$ gives the bracket $[\ ,\
]$. $G^*$ is the dual Poisson-Lie group of $G$. The two Poisson brackets on $G$, $G^*$, which are dually related to the Lie algebra structure on ${\gl g}^*, 
 {\gl g}$, respectively, when evaluated at the identity of the group are nothing but the Kirillov-Souriau-Konstant brackets on  coadjoint orbits of Lie groups. 
The Lie group $D$, associated with the Lie algebra ${\gl d}= {\gl g}\bowtie {\gl g}^*$
is the Drinfel'd double group of $G$ (or $G^*$, being the construction symmetric).\footnote{Properly \cite{semenov}  Drinfel'd doubles are  the quantum version of  double groups. The latter is introduced below. 
The notation classical and quantum Drinfel'd doubles is also used. }\footnote{We denote with the symbol $\bowtie$ the Lie algebra structure of $\mathfrak{d}$ which is totally noncommutative, being both Lie subalgebras non-Abelian. }   

There is a dual  algebraic approach to the picture above, mainly due to Drinfel'd \cite{drinfel'd}, which starts from a deformation of the semi-direct sum $\mathfrak{g}~\dot\oplus ~\R^n$, with $\R^n \simeq\mathfrak{g}^*$, into a fully non-Abelian Lie algebra, which coincides with $\mathfrak{d}$. The latter construction is reviewed below.

To be specific to our problem, we focus on the group $SU(2)$ whose Drinfel'd double can be seen to be the group $SL(2,\C)$ \cite{drinfel'd}. } An action can be shown to be written on the tangent bundle of  $D$, in such a way that the usual Lagrangian description of the rotator can be recovered by reducing the carrier manifold to the tangent bundle of $SU(2)$. 
\label{sl2}

The structure of $\mathfrak{d}=\mathfrak{sl}(2,\mathbb{C})$ as a double algebra is shortly reviewed here.

With this purpose, we start recalling that the complex Lie algebra $\mathfrak{sl}(2)$ is completely defined by the Lie brackets of its generators:
\begin{equation}
[t_3,t_1]=2t_1; \quad [t_3,t_2]=-2t_2; \quad [t_1,t_2]=t_3;  
\end{equation}
with
\begin{equation}
t_1=
\begin{pmatrix}
0 & 1 \\
0 & 0
\end{pmatrix}
; \quad t_2=
\begin{pmatrix}
0 & 0 \\
1 & 0
\end{pmatrix}
; \quad t_3=
\begin{pmatrix}
1 & 0 \\
0 & -1
\end{pmatrix}
.
\label{gensl}
\end{equation}

By considering  complex linear combinations of the basis elements of $\mathfrak{sl}(2)$, say $e_i$, $b_i$,  $i=1,2,3$, respectively given by:
\begin{equation} \label{b1}
e_1  =\frac{1}{2}(t_1+t_2)=\frac{\sigma_1}{2}, \;  \;\; e_2  =\frac{i}{2}(t_2-t_1)=\frac{\sigma_2}{2}, \;\;\;
e_3  =\frac{1}{2}t_3=\frac{ \sigma_3}{2}
\ee
\be
b_i= i e_i   \;\;\; i=1,2,3
\ee
 the real algebra $\mathfrak{sl}(2,C)$ can be easily obtained with its Lie brackets:
 \beqa 
  [e_i,e_j] &=& i{\epsilon_{ij}}^k e_k \label{su} \\
  {[}e_i,b_j{]}&=& i{\epsilon_{ij}}^kb_k \\
 {[}b_i,b_j{]}&=&-i{\epsilon_{ij}}^ke_k 
 \eeqa
 with $\{e_i\},  i=1,2,3$, generating the $\mathfrak{su}(2)$ subalgebra.  
  
In a similar way, one can introduce the combinations:
\begin{equation}
\tilde e^1=it_1;\qquad \tilde e^2=t_1; \qquad \tilde e^3=\frac{i}{2} t_3,
\label{basisdouble}
\end{equation}
which are the dual basis of the generators \eqref{b1},  with respect to the scalar product naturally defined on $\mathfrak{sl}(2,\mathbb{C})$ as:
\begin{equation}
\Braket{u,v}=2\, {\rm Im}(\,Tr(uv)\,), \quad \forall u,v \in \mathfrak{sl}(2,\mathbb{C}).
\label{psd}
\end{equation}
Indeed, it is  easy to show that 
\begin{equation}
\Braket{\tilde e^i, e_j}=2\, {\rm Im}(\,Tr(\tilde e^i e_j)\,)=\delta^i_j  \label{eupedown}.
\end{equation}
Hence, $\{\tilde e^j\}$ is the dual basis of $\{e_i\}$ in the dual vector space $\mathfrak{su}(2)^*$.
Such a vector space is in turn a Lie algebra, the  special Borel subalgebra $\mathfrak{sb}(2,\mathbb{C})$  with the following Lie brackets:
\begin{equation}
[\tilde e^1,\tilde e^2]=0; \qquad [\tilde e^1,\tilde e^3]=- i \tilde e^1; \qquad [\tilde e^2,\tilde e^3]=-i \tilde e^2.  
\end{equation}
  In a more compact form, the generators \eqref{basisdouble} can be written as:
\begin{equation}
\tilde e^i=\delta^{ij}(b_j+e_k{\epsilon^k}_{j3}),
 \end{equation}
and the corresponding the Lie brackets can be derived:
\begin{equation}
[\tilde e^i, \tilde e^j]= i {f^{ij }}_k \tilde e^k \label{sb}
\ee
 and
\begin{equation}
[\tilde e^i,e_j]= i\epsilon^i_{\,jk}\tilde e^k+ i e_k {f^{ki}}_j  
\label{liemis}
\end{equation}
with ${f^{ij}}_k=\epsilon^{ij l}\epsilon_{l3k}$. For future convenience we also note that:
\be
\tilde e^i \te^j= -\frac{1}{4} \delta^{i3}\delta^{j3}\sigma_0 +\frac{i}{2} {f^{ij}}_k \te^k.\label{ee}
\ee
The following relations can be easily checked:
\begin{equation}
\Braket{e_i,e_j}=\Braket{\tilde e^i,\tilde e^j}=0
\ee
so that both $\mathfrak{su}(2)$ and  $\mathfrak{sb}(2,\mathbb{C})$ are maximal isotropic subspaces of $\mathfrak{sl} (2,\mathbb{C})$   with respect to the scalar product \eqref{psd}.\footnote{Notice that another splitting of the $\mathfrak{sl} (2,\mathbb{C})$ Lie algebra into maximally isotropic subspaces with respect to the same scalar product is represented by the span of $\{e_i\}, \{b_i\}, i=1,2,3$, with $\Braket{e_i, e_j}=\Braket{b_i,b_j}=0, \Braket{e_i,b_j}=\delta_{ij}$. However the generators  $\{b_i\}$ do not close a Lie subalgebra.} 
Therefore,  the Lie algebra $\mathfrak{sl}(2,\mathbb{C})$ can be split into two maximally isotropic dual Lie subalgebras with respect to a bilinear, symmetric, non degenerate form defined on it.  The couple   ($\mathfrak{su}(2)$, $\mathfrak{sb}(2,\C)$), with the dual structure described above, is a Lie bialgebra. Since the role of $\mathfrak{su}(2)$ and its dual algebra can be interchanged,  $(\mathfrak{sb}(2,\C)$,  $\mathfrak{su}(2)$) is a Lie bialgebra as well.  The triple  $(\mathfrak{sl}(2,\mathbb{C}), \mathfrak{su}(2), \mathfrak{sb}(2,\mathbb{C}))$ is called a  {\it Manin triple} \cite{drinfel'd}. The total algebra $\mathfrak{d}= \mathfrak{g}\bowtie \mathfrak{g}^*$ which is the Lie algebra defined by the Lie brackets \eqn{su}, \eqn{sb}, \eqn{liemis}, with its dual $\mathfrak{d}^*$ is also  a Lie  bialgebra.

The couple $(\mathfrak{d},\mathfrak{d}^*)$  is called the {\em double} of $(\mathfrak{g},\mathfrak{g}^*)$ \cite{semenov}.  The {\em double group} $D$ is meant to be the Lie group of $\mathfrak{d}$ endowed with some additional structures such as a Poisson structure on the group manifold compatible with the group structure; more details are given in the next section. The two partner groups, $SU(2)$ and $SB(2,\mathbb{C})$ with suitable Poisson brackets,  are named  {\it dual groups} and  sometimes indicated by $G, G^*$. Their role can be interchanged, so that they  share the same double group $D$.  

The splitting of $\mathfrak{sl}(2,\mathbb{C})$ is realized with respect to the scalar product \eqref{psd}.
This  is given by the {Cartan-Killing metric} $g_{ij}=\frac{1}{2}{c_{ip}}^q {c_{jq}}^p$, induced by the structure constants ${c_{ij}}^k$ of $\mathfrak{sl}(2,\mathbb{C})$ in its adjoint representation. 

But  this is not the only decomposition of $\mathfrak{sl}(2,\mathbb{C})$ one can give. There is another non-degenerate, invariant scalar product,  represented by
\begin{equation}
(u,v)=2 Re( \, Tr(u v) \,) \qquad \forall u,v \in \mathfrak{sl}(2,\mathbb{C}). 
\label{sp2}
\end{equation}
In this case, for the basis elements, one gets:
\begin{equation}
(e_i,e_j)=\delta_{ij}, \quad (b_i,b_j)=-\delta_{ij}, \quad (e_i,b_j)=0,
\label{otherscalar}
\end{equation}
giving rise to a metric which is  not positive-definite. With respect to the scalar product defined in eq. (\ref{sp2}), new maximal isotropic subspaces  can be defined in terms of: 
\be
f_i^+=\frac{1}{\sqrt 2} (e_i+ b_i) \,\,\, \quad ; \,\,\, \quad f_i^-= \frac{1}{\sqrt 2}(e_i-b_i) \label{newb}  \,\,\,.
\ee
It turns out that:
\begin{equation}
(f^+_i,f^+_j)= (f^-_i,f^-_j)=\,\,0 \quad ;  \quad (f^+_i,f^-_j)=\delta_{ij} 
\end{equation}
whereas 
\begin{equation}
\Braket{f^+_i,f^+_j}= \delta_{ij}, \quad \Braket{f^-_i,f^-_j}=-\delta_{ij}, \quad \Braket{f^+_i,f^-_j}=0\,\,.  
\end{equation}
Let us notice that neither of them spans a Lie subalgebra. By denoting by $C_+$ and $ C_-$ the two  subspaces spanned by $\{e_i\}$ and $\{b_i\}$ respectively,  one can notice \cite{gualtieri:tesi} that the splitting ${\mathfrak d}= C_{+} \oplus C_{-}$ defines a positive definite metric $\mathcal{H}$ on ${\mathfrak d}$ via:
\be
\mathcal{H}= (\;,\;)_{C_+}-  (\;,\;)_{C_-} \label{metricG}
\ee
As in ref. \cite{gualtieri:tesi},  the inner product is here used to identify ${\mathfrak d}$  with its dual, so that the metric $\mathcal{H}$ may be viewed as an
automorphism of ${\mathfrak d}$ which is symmetric and which squares to the identity, i.e.
$\mathcal{H}^2 = 1$. Let us indicate the Riemannian metric with double round brackets. One has then:
\be
((e_i,e_j)) \equiv (e_i,e_j); ~~~~~ ((b_i,b_j)) \equiv -(b_i,b_j);~~~~~ ((e_i,b_j)) \equiv (e_i,b_j)=0  \,. \label{riem}
\ee

In order to come back to the main subject of the paper, namely the relation between GG and DFT, introducing the following notation for the $\mathfrak{sl}(2,\mathbb{C})$ generators reveals to be very helpful:
%=\mathfrak{su}(2) \oplus \mathfrak{sb}(2,\mathbb{C})$, we can write the orthogonal elements of 
%$\mathfrak{sl}(2,\mathbb{C})$ as
\begin{equation}
e_I=\begin{pmatrix}e_i\\ e^i \end{pmatrix},
 \qquad  e_i \in \mathfrak{su}(2), \quad  e^i \in \mathfrak{sb}(2,\mathbb{C}),
\label{doubledb}
\end{equation}
with $I=1, \dots 2d$, being $d= {\rm dim} \,\mathfrak{g}$. Then  the scalar product \eqn{psd} becomes
\begin{equation} \label{Lprod}
\Braket{e_I,e_J}={\cal \eta}_{IJ}= 
\begin{pmatrix}
0 & \delta_i^j \\
\delta_j^i &  0
\end{pmatrix}
. 
\end{equation}
{This symmetric inner product has signature  $(d,d)$ and therefore defines the non-compact orthogonal group  $O(d,d)$, with $d=3$ in this case}.

The Riemannian product \eqn{riem} yields instead:
\beqa
((\tilde e^i, \tilde e^j))&=&\delta^{ip}\delta^{jq} ((b_p + e_l{\epsilon^l}_{p3} ))((b_q+ e_k{\epsilon^k}_{j3} )) \nonumber \\
&=& \delta^{ij}+\epsilon^i_{\;l3} \delta^{lk}  \epsilon^j_{\;k3} \,\,\,\,\, ; \label{sp2ee}\\
(( e_i, \tilde e^j))&=& [(e_i, b_q) + {\epsilon^k}_{q3} (e_i, e_k)] \delta^{jq}={\epsilon_{3i}}^j \,\,\, .\label{mixed}
\eeqa
Hence,  one has:
\begin{equation} \label{Rprod}
((e_I,e_J))={\cal H}_{IJ}= 
\begin{pmatrix}
\delta_{ij} & {\epsilon_{3i}}^j \\
-{\epsilon^{i}}_{j3} &  \delta^{ij}+ \epsilon^i_{\;l3} \delta^{lk}\epsilon^j_{\;k3} 
\end{pmatrix}
. 
\end{equation}
This metric satisfies the relation: 
\be {\cal H}^T {\cal \eta} {\cal H}= \eta \label{compatibile}
\ee
indicating that ${\cal H}$ is a pseudo-orthogonal $O(3,3)$ matrix.

It is interesting to see how  the metric ${\cal \eta}$ in eq. (\ref{Lprod})  and the metric ${\cal H}$ in eq. (\ref{Rprod}) naturally emerge out in the framework here in exam. They correspond, in the usual context of DFT, to the $O(d,d)$ invariant metric and to the so-called  {\em generalized metric} \cite{Tseytlin, Hohm}, respectively.  In particular, in the latter, the role of the graviton field is played by the Kronecker delta $\delta_{ij} $ while the role of the Kalb-Ramond field  is played by the three-dimensional Levi-Civita symbol $\epsilon_{ij3}$ with one of the indices being fixed.

\subsection{Para-Hermitian Geometry of $SL(2,\mathbb{C})$}

The two non-degenerate scalar products of $SL(2,\C)$, discussed above, have been widely applied in many physical contexts where the Lorentz group and its universal covering $SL(2,\C)$ play a role, starting from the pioneering work by E. Witten \cite{witten}. 
While the first scalar product, i.e. the one defined in eq. (\ref{psd}), is nothing but the Cartan-Killing metric of the algebra, the Riemannian structure ${\cal H}$ can  be mathematically formalized in  a way which clarifies its role in the context of Generalized Complex Geometry  \cite{gualtieri:tesi, freidel} and gives a further example of doubled geometry \cite{hulled}. Let us shortly review the derivation.

The splitting of $\mathfrak{sl}(2,\mathbb{C})$ in $\mathfrak{su}(2)$ and $\mathfrak{sb}(2,\mathbb{C})$ implies the existence of a $(1,1)$-tensor:
\be \mathcal{R}: \mathfrak{sl}(2,\mathbb{C})  \rightarrow \mathfrak{sl}(2,\mathbb{C}) \label{Rtensor}
\ee
such that $\mathcal{R}^2=\mathds{1}$ and of eigenspaces given by $\mathfrak{su}(2)$, with eigenvalue $+1$, and $\mathfrak{sb}(2,\mathbb{C})$, with eigenvalue $-1$. This can be seen as the local expression of  a $(1,1)$-tensor on $SL(2,\mathbb{C})$ called \emph{product structure}, since it has integrable eigenbundles $TSU(2)$ and $TSB(2,\mathbb{C})$, that, at every point of $SL(2,\mathbb{C})$, are given by $\mathfrak{su}(2)$ and $\mathfrak{sb}(2,\mathbb{C})$ and are such that $TSL(2,\mathbb{C})=TSU(2) \oplus TSB(2,\mathbb{C}).$
These two eigenbundles are maximal isotropic with respect to the scalar product \eqn{psd} and, being integrable, they give rise to two transversal foliations of $SL(2,\mathbb{C})$, the one  with $SU(2)$ as leaves, the other with $SB(2,\mathbb{C})$. 

Moreover, the tensor $\mathcal{R}$ is compatible with the scalar product \eqn{psd} meaning that the following equation holds:
\be \Braket{\mathcal{R}(X),Y}+\Braket{\mathcal{R}(Y),X}=0 \quad \forall X,Y \in \Gamma (TSL(2,\mathbb{C})),  \nonumber
\ee
where $\Gamma (TSL(2,\mathbb{C}))$ denotes the vector fields over the group manifold. In a more compact form, one has: $\mathcal{R}^T {\cal \eta} \mathcal{R}=-L$, with ${\cal \eta}(X,Y)=\Braket{X,Y},\ \forall X,Y \in \Gamma (TSL(2,\mathbb{C}))$.
This condition implies that a $2$-form field can be defined as: $$\omega(X,Y)= \Braket{\mathcal{R}(X),Y}, \quad \forall X,Y \in \Gamma (TSL(2,\mathbb{C})).$$
In other words, a \emph{para-Hermitian structure} \cite{freidel}  can be defined on the manifold $SL(2,\mathbb{C})$, where $\mathcal{R}$ is the product structure,  \eqn{psd} is the scalar product compatible with the Lorentzian signature and $\omega$ is the fundamental two-form.
In this sense, the scalar product \eqn{Rprod} can be read as a metric with Riemannian signature considering the bases \eqn{newb}. 

In fact, expressing the bases \eqn{newb}  as linear combinations of  $\{e_i\}$ and $\{\tilde{e}^i\}$ yields the following: 
\be f^+_i= \frac{1}{\sqrt{2}}\bigl( \delta_{ij}\tilde{e}^j+(\delta^j_i+{\epsilon_{3i }}^{j})e_j \bigr), \label{baseup}
\ee
and
\be f^-_i= \frac{1}{\sqrt{2}}\bigl( -\delta_{ij}\tilde{e}^j+(\delta^j_i+{\epsilon_{3i }}^{j})e_j \bigr), \label{baseum}
\ee
which  generate, respectively, the subspaces $V_+$ and $V_-$ of $\mathfrak{sl}(2,\mathbb{C})$, maximal isotropic with respect to \eqn{sp2}.

Then, given the splitting $\mathfrak{sl}(2,\mathbb{C})=V_+ \oplus V_-$,  there exists the $(1,1)$-tensor $H$ such that
\be H (f^+_i)=f^+_i, \quad H(f^-_i)=-f^-_i, \label{locale}
\ee
and $$H^2=\mathds{1},$$ implying  that $V_+$ and $V_-$ are eigenspaces of  $H$ with eigenvalues $+1$ and $-1$ respectively.

One can immediately notice that the Lie bracket of any two elements in $\{f^+_i\}$ and $\{f^-_i\}$ is not involutive, hence $V_+$ and $V_-$ are not Lie subalgebras of $\mathfrak{sl}(2,\mathbb{C})$. 

As described at the beginning of this section, eq. \eqn{locale} can be read as the definition, at any point, of a $(1,1)$-tensor field $H$ as an \emph{almost product structure} on $SL(2,\mathbb{C})$, since the eigenbundles $\mathcal{V}_+$ and $\mathcal{V}_-$, obtained as distributions that, at any point, are $V_+$ and $V_-$ respectively, are not integrable. The local change of splitting implies that $TSL(2,\mathbb{C})=\mathcal{V}_+ \oplus \mathcal{V}_-$.

In order to write down the dual bases $\{f^{j*}_+\}$ and $\{f^{j*}_-\}$ of $\{f^+_j\}$ and $\{f^-_j\}$ respectively, by using the duality relation between $\{e_i\}$ and $\{\tilde{e}^i\}$, the duality conditions have to be imposed:
\be  f^+_i(f^{j*}_+)=\delta_i^j, \quad f^-_i(f^{j*}_+)=0 \nonumber
\ee
and
\be  f^-_i(f^{j*}_-)=\delta_i^j, \quad f^+_i(f^{j*}_-)=0 \nonumber
\ee
which lead to 
\be f^{i*}_+=\frac{1}{\sqrt{2}}\bigl(\tilde{e}^i+(\delta^{ik}+\epsilon^{ik3})e_k\bigr)  \nonumber
\ee
and
\be f^{i*}_-=\frac{1}{\sqrt{2}}\bigl(\tilde{e}^i-(\delta^{ik}+\epsilon^{ki3})e_k\bigr).   \nonumber
\ee
Therefore,  the almost product structure $H$ turns out to be the following:
\be H=\delta^i_j f^+_i \otimes f_+^{j*} - \delta^i_j f^-_i \otimes f_-^{j*},  \nonumber
\ee
which, in the bases $\{e_i\}, \ \{\tilde{e}^i\}$, becomes:
\be H=\delta_{ij}\tilde{e}^i \otimes \tilde{e}^j+ \delta_{ik}\epsilon^{kj3}\tilde{e}^i \otimes e_j + \delta_{kj}\epsilon^{ki3}e_i \otimes \tilde{e}^j + (\delta^{ij}+ \delta_{lk}\epsilon^{il3}\epsilon^{jk3})e_i \otimes e_j .   \nonumber
\ee
The metric \eqn{Lprod} can be used on $H$ 
for raising and lowering indices. In fact, in the doubled formalism, one can write the matrix:
\begin{equation} 
{\cal H}_{IJ}={\cal \eta}_{IK} H^K_{\ J}=
\begin{pmatrix}
\delta_{ij} & {\epsilon_{i}}^{j3} \\
-{\epsilon^{i}}_{j3} &  \delta^{ij}+\delta_{lk}\epsilon^{il3}\epsilon^{jk3}
\end{pmatrix}
\label{splitpos}
\end{equation}
which is exactly the generalized metric \eqn{Rprod}, i.e. $\mathcal{H}= \eta H$. The metric \eqn{splitpos} is a representative element in the coset $O(3,3)/O(3) \times O(3),$ i.e. it is defined by $3^2$ independent elements. Thus, the form of the generalized metric depends on the choice of polarization (the splitting of the double Lie algebra in two maximal isotropic subspaces). In fact, as in Generalized Geometry, the metric \eqn{splitpos} gives a reduction of the structure group of $TSL(2,\mathbb{C})$ to $O(3)\times O(3)$ and we can interpret such reduction, in this very specific case, as related to the other natural scalar product on $SL(2,\mathbb{C}).$ Moreover, the introduction of a generalized metric on Drinfel'd doubles is also discussed in \cite{hulled}, in which has been made clear how, from different choices of polarization of the Drinfel'd double, the generalized metric takes a different form which allows to recover different backgrounds. This is shown to be very useful in the description and gauging of non-linear sigma models with doubled target space.

It is also easy to verify that the metric with Lorentzian signature arising from \eqn{sp2} is given by: 
\be
K=\delta_{ij} f_+^{i*} \otimes f_-^{j*} + \delta_{ij} f_-^{i*} \otimes f_+^{j*}
\ee which takes the form 
\begin{equation} 
K_{IJ}=
\begin{pmatrix}
\delta_{ij} & {\epsilon_{i}}^{j3} \\
-{\epsilon^{i}}_{j3} &  -\delta^{ij}+\delta_{lk}\epsilon^{il3}\epsilon^{jk3}
\end{pmatrix}   \nonumber
\end{equation}
 and is compatible with $H$, i.e. $H^T K H=-K.$
The compatibility condition of $K$ and $H$ gives a closed two-form 
\be
\Omega = K H= \delta_{ij} f^{i*}_+ \wedge f^{j*}_-.
\ee
 that, in  the bases  $\{e_i\}$ and $\{\tilde{e}^i\}$, gets the following expression:
 \be
 \Omega=\delta^i_j e_i \otimes \tilde{e}^j - \delta_i^j \tilde{e}^i \otimes e_j .  \nonumber
 \ee
This can be read as an explicit form of the product structure $\mathcal{R}$ and the fundamental two-form $\Omega$.

In conclusion,  it has been here shown that the natural scalar product \eqn{sp2} and the almost product structure $H$ define an \emph{almost para-Hermitian structure} on the manifold $SL(2,\mathbb{C}).$

Finally, let us describe the structure arising from \eqn{riem}. In the previous section, a 
 positive definite metric $\mathcal{H}$  \eqn{metricG} on $SL(2,\mathbb{C})$ has been defined by the splitting $\mathfrak{sl}(2,\mathbb{C})=C_+ \oplus C_-.$
In order to explicitly write down the metric tensor $\mathcal{H}$,  dual bases of $\{ e_i\}$ and $\{ b_i \}$ have been introduced.
After noticing that $$b_i=\delta_{ij} \tilde{e}^j +\epsilon_i^{\ k3}e_k$$ and by imposing the conditions $$ e_i (b^{*j})=0\,\,\,\,\,, \ \ \ b_i (b^{j*})=\delta_i^j$$ and $$e_i (e^{j*})=\delta_i^j\,\,\,\,\,, \ \ \ b_i (e^{j*})=0\,\, , $$
one obtains: $$e^{i*}= \tilde{e}^i + \epsilon^{ik3}e_k\,\,\,\,\,, \qquad b^{i*}=\delta^{ij}e_j.$$ It is worth to stress that changing the splitting also changes the dual bases.
Thus, the metric tensor of eq. \eqn{metricG} can be retrieved:
\be 
\mathcal{H}=\delta_{ij}e^{i*} \otimes e^{j*}+\delta_{ij}b^{i*} \otimes b^{j*}=(,)_{C_+}-(,)_{C_-}\,\,\,.  \nonumber
\ee
It takes the form \eqn{splitpos} in the bases ($\{e_i\}, \ \{\tilde{e}^i\}$), is symmetric and squares to the identity.
Moreover, the metric $\mathcal{H}$ can be seen to be  given by the composition of two {\it generalized complex structures} \cite{gualtieri:tesi}, $\mathcal{I}_J$ and $\mathcal{I}_{\omega}$, respectively defined by an almost complex structure $J$ and a symplectic structure $\omega$.  Therefore, one has  a pair $(\mathcal{I}_J,\mathcal{I}_{\omega})$ of commuting generalized complex structures on $TSL(2,\mathbb{C})$ inducing a positive definite metric $\mathcal{H}$. This is usually called a \emph{generalized K\"ahler structure}.
Consequently,  it has been shown how the almost para-Hermitian structure and the generalized metric $\mathcal{H}$ on the manifold $SL(2,\mathbb{C})$ are related. 

The discussion that has been just completed shows the existence of two non-degenerate, invariant scalar products on the  total algebra $\mathfrak{d}$. 
 In the forthcoming sections  both products will be considered in order to define action functionals for the dynamical systems in exam.

 \section{The Dual Model}\label{dualrot}
In the previous section the dual group of $SU(2)$, the group $SB(2,\C)$, has been introduced as the partner of $SU(2)$ in a kind  of  Iwasawa decomposition of the group $SL(2,\C)$. The latter has been regarded as a deformation of the cotangent bundle of $SU(2)$ with  fibers $F\simeq \R^3$ replaced by  the group $SB(2,\C)$. It is then legitimate to reverse the paradigma and regard $SL(2,\C)$ as a deformation of the cotangent bundle $T^*SB(2,\C)$, with  fibers $\tilde F\simeq\R^3$ now replaced by $SU(2)$. 
In this section a dynamical model on the configuration space $SB(2,\C)$  is proposed with an action functional that is  formally analogous and indeed dual to \eqn{lag}. The model is described with its symmetries in the Lagrangian and Hamiltonian formalisms. 

In Sect. \ref{gensec} a generalized action containing both the models is finally introduced on the whole group $SL(2,\C)$. The Poisson algebra encoding the dynamics, as well as the algebra of generalized vector fields describing infinitesimal symmetries, turns out to be related to the so-called $C$-bracket of DFT.

\subsection{The Lagrangian and Hamiltonian Formalisms}
 As carrier space for the dynamics of the dual model in the Lagrangian (respectively Hamiltonian) formulation one can choose  the tangent (respectively cotangent) bundle of the group $SB(2,\C)$. 
A suitable action for the system  is the following:
\be
{\tilde S}_0= \int_\R {\tilde L} _0~dt= - \frac{1}{4} \int_\R {\mathcal Tr} [\tg^{-1} d \tg\wedge * \tg^{-1} d\tg] = - \frac{1}{4}\int_\R   {\mathcal Tr} [(\tg^{-1}{ \dot \tg})(\tg^{-1}{ \dot \tg})] dt \label{dualag}
\ee
with $\tg:t\in \R\rightarrow SB(2,\C)$, the group-valued target space coordinates, so that 
\be
\tg^{-1} d \tg= i \beta_k \te^k  \nonumber
\ee
 is the Maurer-Cartan left invariant one-form on the group manifold, with $\beta_k$ the left-invariant basic one-forms,   $*$ the Hodge  
star operator on the source space $\R$, such that $* dt = 1$. The symbol ${\mathcal Tr} $ is here used to represent a {\it suitable} scalar product in the Lie algebra $\mathfrak{sb}(2,\C)$. Indeed, since the algebra is not semi-simple, there is no scalar product which is both non-degenerate and invariant. Therefore,  one has two possible different choices: the scalar product defined by the real and/or imaginary part of the trace, given by Eqs. \eqn{psd} and \eqn{sp2}  which is   $SU(2)$ and $SB(2,\C)$ invariant, but degenerate; or one could use the scalar product induced by the Riemannian metric $G$, which, on the algebra $\mathfrak{sb}(2,\C)$ takes the form  \eqn{sp2ee} which is positive definite and non-degenerate, but only invariant under left $SB(2,\C)$ action and $SU(2)$ invariant.
Indeed, by observing that the generators $\te^i$ are not Hermitian, \eqn{sp2ee} can be verified to be equivalent to:
\vspace{0.3cm}
\be
((u,v)) \equiv 2{\rm Re}\Tr [(u)^\dag v]  \label{2ndprod}
\ee
so that $((\tg^{-1}{ \dot \tg}, \tg^{-1}{ \dot \tg}))= 2{\rm Re}\Tr [(\tg^{-1}{ \dot \tg})^\dag \tg^{-1}{ \dot \tg}]$ which 
is not invariant under right $SB(2,\C)$ action, since  $\tg^{-1}\ne \tg^\dag$. 

 The associated dynamical models are obviously different.   The non-degenerate scalar product defined  in eq. \eqn{sp2ee} is used here, therefore the Lagrangian \eqn{dualag} is only left/right $SU(2)$ and left-$SB(2,\C)$ invariant, differently from  the Lagrangian of the rigid rotator \eqn{lag} which is invariant under both left and right actions of both groups.  As in the previous case, the model can be regarded as a $(0+1)$-dimensional field theory which is 
 group-valued. 
 
The group manifold can be parametrized with  $\R^4$ coordinates, so that  
$\tg\in SB(2,\C)$ reads $\tg= 2( u_0 \te^0 
+ i  u_i \te^i)$, with $u_0^2- u_3^2=1$ and $\te^0= {\mathbb I}/2$.  One has then:
\be
u_i=\frac{1}{4} (( i\tg, \te^i)), \;i=1,2,~~~~ \; u_3=\frac{1}{2} (( i\tg, \te^3)), ~~~~\; u_0= \frac{1}{2}((\tg,  \te^0))  \nonumber \;\;\;
\ee
where the last product is defined as twice the real part of the trace, in order to be consistent with the others.
By observing that 
\be
\tg^{-1} \dot \tg=2 i  (u_0\dot u_i-u_i \dot u_0+  {f_{i}}^{\,jk} {u_j \dot u_k })\te^i  \label{tiqdot}
\ee
the Lagrangian in \eqn{dualag} can be rewritten as:
\be
\tilde{{L}}_0= (u_0\dot u_i-u_i \dot u_0+  {f_{i}}^{\,jk} {u_j \dot u_k })(u_0\dot u_r-u_r \dot u_0+  {f_{r}}^{\,pq} {u_p \dot u_q })((\te^i,\te^r)) 
= \dot \tQ_i \dot \tQ_r  h^{ir}   \nonumber
\ee
being
\be
\dot \tQ_i \equiv u_0\dot u_i-u_i \dot u_0+  {f_{i}}^{\,jk} {u_j \dot u_k }  \nonumber
\ee
the left generalized velocities
and
\be
h^{ir} \equiv (\delta^{i r}+ {\epsilon^i}_{l3}{\epsilon^r}_{s3}\delta^{ls}) \label{hir}
\ee
the metric defined by the scalar product. By repeating the analysis already performed for the IRR,  one finds the equations of motion:
\be
{\sf L}_\Gamma(\dot \tQ_j\, i_{\tX^i}\beta_l )h^{jl} - {\sf L}_{\tX^i} \tilde{L}_0 = 0.  \label{LGamma}
\ee
with $\tX^j$ being the left invariant vector fields associated with $SB(2,\C)$. Differently from the IRR case, the Lagrangian now is not invariant under right action, therefore, being the left invariant vector fields the generators of the right action, the l.h.s.  in eq. (\ref{LGamma}) is not expected to be zero and through a straightforward calculation it results to be:
\be
{\sf L}_\Gamma\dot \tQ_j h^{ji} - \dot \tQ_p \dot \tQ_q {f^{ip}}_k h^{qk} =  0. \label{eqmodu}
\ee
$(\tQ_i, \dot \tQ_i)$ are therefore  tangent bundle coordinates, with $\tQ_i$ implicitly defined. 

It has to be noticed here that, analogously to the IRR case, one could  define the  right generalized velocities on the fibers starting from right invariant one-forms, but, differently from that case, the right invariant Lagrangian is not equivalent to the left invariant one, as already stressed.  

 The cotangent bundle coordinates are $(\tQ_i, \tI^i)$ with $\tI^i$ the conjugate left momenta 
\be
\tI^j= \frac{\del {\tilde{{ L}}}_0}{\del \dot \tQ_j}= \dot \tQ_r (\delta^{j r}+ \epsilon^j_{\,l3}\epsilon^r_{s3}\delta^{ls}) = \frac{i}{2} ((\tg^{-1}\dot\tg,\te^i))\delta_{i}^j \,\,\, . \nonumber
\ee
The latter is in turn invertible, yielding:
\be
\dot\tQ_j= \tI^i(\delta_{ij}-\frac{1}{2}\epsilon_{ip3}\epsilon_{jq3}\delta^{pq}),  \nonumber
\ee
so that 
the Legendre transform from $TSB(2,\C)$ to $T^*SB(2,\C)$ leads to  the Hamiltonian function: 
\be
\tilde{H}_0=[\tI^j \dot \tQ_j -\tilde L]_{\dot \tQ=\dot \tQ(\tI)}= \frac{1}{2}\tI^i (h^{-1})_{ij }\tI^j \,\,\, ,\label{h0du}
\ee
being
\be
 (h^{-1})_{ij } \equiv (\delta_{ij}-\frac{1}{2}\epsilon_i^{\,p3}\epsilon_j^{\,q3}\delta_{pq})
 \label{hinvij}
 \ee
the inverse of eq. \eqn{hir}. 
Similarly to eq.  \eqn{Iform},  the linear combination over the dual basis is introduced:
\be
\tilde I= i \tilde I^j {\te_j}^* \label{tIform}
\ee
with $\langle {e_j}^* |\te^i\rangle=\delta_j^i$.
Then, the first order dynamics can be obtained from the Hamiltonian \eqn{h0du} and the  following  Poisson brackets:
\beqa
\{u_i,u_j\}&=&0\label{ppdu}\\
\{\tI^i,\tI^j\}&=& {f^{ij\;}}_k\tI^k \label{xxdu}\\
\{u_i,\tI^j\}&=&{\delta_{i}^j u_0 -{f_i}^{\; jk}u_k \;\;\; {\rm or ~ equivalently}\;\;\; \{\tg, \tI^j\}= 2 i  \tg \te^j }\label{ijdu}
 \eeqa
which are derived from the first order formulation of the action functional. Since the results are slightly different from the IRR case, let us  present the derivation in some detail. 

The first-order action functional reads in this case as:
 \be
 \tilde S_1= \int \langle \tI|  \tg^{-1}d \tg\rangle  - \int \tilde H dt \equiv \int \tilde \vartheta -\int \tilde H dt \, .
 \ee
Observing that
\be
\langle \tI|  \tg^{-1}d \tg\rangle =i \tI^i \delta^k_i \beta_k \,\, ,
\ee
the symplectic form $\tilde \omega$ reads as:
 \be
 \tilde \omega= d \tilde \vartheta= d \tI^j \wedge \beta_j   - \frac{1}{2}\tI^j {f_j}^{lm} \beta_l\wedge \beta_m \label{tomega}
 \ee
 where the relation 
   $d [\tg^{-1}d \tg] = \frac{i}{2} \beta_i\wedge\beta_j { f^{ij}}_k\te^k$ has been used. By inverting $\tilde \omega$, one finally finds the Poisson algebra \eqn{ppdu}-\eqn{ijdu}.

 Hamilton equations are readily obtained from the Poisson brackets. In particular one gets:
 \be
 \dot\tI^j= \{\tI^j,\tilde H\}= f^{jk}_l\tI^l\tI^r h^{-1}_{kr}   \nonumber
 \ee
 which is consistent with  eq. \eqn{eqmodu} and is different from  zero, expressing the non-invariance of the Hamiltonian under right action. 
 Vice-versa, by introducing the right momenta $\tJ^i$ as the Hamiltonian functions of right-invariant vector fields, which in turn generate the left action, and observing that left and right invariant vector fields commute, 
 one readily obtains:
 \be
 \dot\tJ^j= \{\tJ^j,\tilde H\}= 0
 \ee
 namely, right momenta are constants of the motion and the Hamiltonian is invariant under left action, as we expected. 

 By using \eqn{ijdu} it is possible to find:
 \be
 \tg^{-1}\dot \tg= 2 i \te^i (h^{-1})_{ij} \tI^j   \nonumber
 \ee
 consistently with eq. \eqn{tiqdot}.
 Right momenta are therefore conserved, as for the rigid rotator, while left momenta are not.

 Let us remark here that, while the fibers of the tangent bundle $TSB(2,\C)$ can be identified, as a vector space,  with the Lie algebra of $SB(2,\C)$, $\mathfrak{sb}(2,\C)\simeq \R^3$, with $\dot \tQ_i$ denoting vector fields components, the fibers of the cotangent bundle $T^*SB(2,\C)$ are isomorphic to the dual Lie algebra $\mathfrak{sb}(2,\C)^*$. As a vector space this is again $\R^3$, but $\tI^j$ are now components of one-forms. This remark will be relevant in the next section where the Abelian structure of $\mathfrak{sb}(2,\C)^*$ is deformed.

As a group, $T^*SB(2,\C)$ is the semi-direct product of $SB(2,\C)$ and the Abelian group $\mathbb{R}^3$, with Lie algebra the semi-direct sum represented by
\beqa
\left[B_i,B_j\right]  &=& i {f_{ij}}^k B_k \label{BB}\\
\left[S_i,S_j\right] &=& 0 \label{SS}\\
\left[B_i,S_j\right] &=&i {f_{ij}}^k S_k. \label{BS}
\eeqa
Then, as before, the  non-trivial Poisson bracket on the fibers of the bundle, \eqn{xxdu},  can be 
understood in terms of the coadjoint action of the group $SB(2,\C)$ on 
$\mathfrak{sb}(2,\C)^*\simeq \mathbb{R}^3$ , i.e. its dual algebra, and it reflects the
non-triviality of the Lie bracket \eqn{BB} with the Lie algebra generators $B_i$ 
identified with  linear functions on the dual algebra.

To summarize the results of this section, the model that has been introduced is dual to the Isotropic Rigid Rotator in the sense that  the configuration space $SB(2,\C)$ is dual, as a group, to $SU(2)$. Moreover, as we shall see, the Poisson brackets of the momenta $I_i, \tI^i$ are dually related. 

In the next section, a generalized action is constructed  on the Drinfel'd double group and it encodes the duality relation between the two models and the global symmetries that have been discussed. 

\section{A New  Formalism for the Isotropic Rotator: the Doubled Formulation}\label{gensec}
In the previous sections, two dynamical models have been introduced with configuration spaces being Lie groups which are dually related. The Poisson algebras for the respective cotangent bundles, $T^*SU(2)$, $T^*SB(2,\C)$, which we restate for convenience in the form:
\beqa
\{g,g\}&=&0,\;\;\;\;\{I_i,I_j\}=  {\epsilon_{ij}}^k I_k , \;\;\;\;\; \{g, I_j\}= 2 i  g e_j \label{sudue}\\
\{\tg,\tg\}&=&0,\;\;\;\;
\{\tI^i,\tI^j\}= {f^{ij}}_k\tI^k ,\;\;\;\;
\{\tg, \tI^j\}= 2 i  \tg \te^j  \,\,\, , \label{sbdue}
 \eeqa
have both the structure of a semi-direct sum dualizing the semi-direct structure of the Lie algebras $\mathfrak{su}(2)\dot\oplus\R^3$ and $\mathfrak{sb}(2,\C) \dot \oplus \R^3$. By identifying the dual algebras $\R^3$, in both cases,  with an Abelian  Lie algebra, we have that each semi-direct sum has the the form \eqn{Liesuma}, with $\R^3$ generators satisfying trivial brackets and with a trivial $ad^*$ action:
\be
 [X+\xi,Y+\zeta]=[X,Y]-ad^{*}_{X}\zeta + ad^{*}_{Y}\xi \label{sumatriv} \,\,\,.
\ee
To this, it is sufficient to expand the group variables, $g,\tg$  
\be
 g\simeq \mathds{1}+i \lambda J^i e_i + O(\lambda^2), \;\;\;\; \tg\simeq \mathds{1}+i  \mu \tilde{J_i} \te^i + O(\mu^2)    \label{groupvariables}
 \ee
and compute the related Poisson brackets   in Eqs. \eqn{sudue}, \eqn{sbdue}
to first order in the parameters. One gets:
\beqa
\{J^i, J^j\}&=& 0,\;\;\;\;\{I_i,I_j\}= {\epsilon_{ij}}^k I_k , \;\;\;\;\;\{J^i, I_j\}= -{\epsilon^{i}}_{jk} J^k \label{suduee}\\
\{\tilde {J}_i, \tilde {J}_j\}&=& 0,\;\;\;\;\{\tI^i,\tI^j\}= {f^{ij}}_k\tI^k ,\;\;\;\;\{\tilde {J}_i, \tI^j\}= - \tilde{J}^k{f_{ki}}^j.\label{sbduee}
\eeqa
The new dynamical variables $J^i, \tilde {J}_i$ will be identified in the forthcoming section with $I^i, \tI_i$ by unifying the cotangent bundles $T^*SU(2), T^*SB(2,\C)$ into the Drinfel'd double $SL(2,\C)$. The brackets \eqn{suduee}, \eqn{sbduee} will then emerge naturally as appropriate limits of  the Poisson-Lie brackets on the dual groups  $G$, $G^*$, when evaluated at the identity of the respective groups as in  eq. \eqn{Liesuma}. 

\subsection{The Lagrangian Formalism}
We are now ready to introduce the new action for the Isotropic Rigid Rotator using the Lagrangian formalism on $TSL(2,\mathbb{C})$.
As in the conventional formulation described above, its description can be read as a $(0+1)$-dimensional field theory which is group-valued, with $g(t)\in SU(2)$ now replaced by $\gamma:t\in \mathbb{R} \rightarrow \gamma(t)\in SL(2,\mathbb{C})$. The left invariant one-form on the group manifold is then:
\begin{equation}
\gamma^{-1} \mathrm{d}\gamma= \gamma^{-1} \dot \gamma\; dt \equiv  \dot {\bf Q}^I e_I \mathrm{d}t \label{gammagamma}
\ee
with $e_I=(e_i, \tilde e^i)$ the $\mathfrak{sl}(2,\C)$ basis introduced in eq. \eqn{doubledb} and  $ \dot {\bf Q}^I$, the left generalized velocities. By defining the decomposition $\dot {\bf Q}^I \equiv( A^i, B_i)$ one has:
\be
\gamma^{-1} \dot \gamma \;dt= (A^i e_i + B_i \tilde e^i) dt  \nonumber
\ee
where, however, both components are tangent bundle coordinates for $SL(2,\C)$\footnote{We could alternatively interpret $(A^i, B_i)$ as fiber coordinates of the generalized bundle $T\oplus T^*$, with base manifold $SU(2)$, so that the model is an instance of both Generalized Geometry and Doubled Geometry.}. By using the scalar product \eqn{psd}, the components of the generalized velocity can be explicitly obtained:
\be
A^i= 2{\rm Im} \Tr ( \gamma^{-1}\dot \gamma \tilde e^i); \;\;\; B_i= 2{\rm Im} \Tr (\gamma^{-1}\dot \gamma e_i).  \nonumber
\ee
Since 
\begin{equation}
*\gamma^{-1} \mathrm{d}\gamma= \dot {\bf Q}^I e_I  \,\, ,  \nonumber
\end{equation}
with the Hodge operator defined as previously, namely $* dt = 1$, 
the proposed action is the following:
\begin{equation}
{S}= \int_R {L} dt=  \frac{1}{2}\int_{\mathbb{R}}\bigl( k_1\Braket{\gamma^{-1}\mathrm{d}\gamma\stackrel{\wedge}{,}* \gamma^{-1}\mathrm{d}\gamma} +k_2  ((\gamma^{-1}\mathrm{d}\gamma \stackrel{\wedge}{,} * \gamma^{-1}\mathrm{d} 
\gamma)) \bigr),
\label{newac}
\end{equation}
where $k_1,k_2$  are  real parameters,  and 
$ \Braket{\gamma^{-1}\mathrm{d}\gamma \stackrel{\wedge}{,} * \gamma^{-1}\mathrm{d}\gamma}$ is defined in terms of the scalar product in eq. \eqn{Lprod} while  $((\gamma^{-1}\mathrm{d}\gamma \stackrel{\wedge}{,} * \gamma^{-1}\mathrm{d}\gamma))$ is defined in terms of the scalar product in eq. \eqn{Rprod}, namely:
\beqa
 \Braket{\gamma^{-1}\mathrm{d}\gamma \stackrel{\wedge}{,} * \gamma^{-1}\mathrm{d}\gamma}
 &=&  \dot {\bf Q}^I  \dot {\bf Q}^J  \Braket {e_I,e_J}=   \dot {\bf Q}^I  \dot {\bf Q}^J  \eta_{IJ}\\
((\gamma^{-1}\mathrm{d}\gamma\stackrel{\wedge}{,} *\gamma^{-1}\mathrm{d}\gamma))&=&  \dot {\bf Q}^I  \dot {\bf Q}^J  ((e_I,e_J))=   \dot {\bf Q}^I  \dot {\bf Q}^J \mathcal{H}_{IJ}.
\label{prodotto2}
\eeqa
Explicitly, in terms of the chosen splitting of the Drinfel'd double $\mathfrak{sl}(2,\mathbb{C})=\mathfrak{su}(2) \bowtie \mathfrak{sb}(2,\mathbb{C})$, one has, up to an overall constant:
\be \label{explito}
{L}=   \frac{1}{2} ( k\,\Braket{e_I,e_J}\dot {\bf Q}^I  \dot {\bf Q}^J  +  (e_I,e_J)\dot {\bf Q}^I  \dot {\bf Q}^J )=       \frac{1}{2}  ( k\, {\cal \eta}_{IJ} +  {\cal H}_{IJ})\dot {\bf Q}^I  \dot {\bf Q}^J  
 \ee
 with 
 \be
k\, {\cal \eta}_{IJ}+  {\cal H}_{IJ}=  
\begin{pmatrix}
\delta_{ij}& k \delta_i^j +    {\epsilon_{3i}}^{j} \\
-{\epsilon^i}_{j3}+k \delta_i^j& \delta^{ij}+ {\epsilon^i}_{l3} \epsilon^j_{k3}\delta^{lk}
\end{pmatrix}   \nonumber
\ee  
and where the  position $k_1/k_2 \equiv  k $ has been made. This leads to:
\be \label{newlag}
{L} =   \frac{1}{2} \bigl[  \delta_{ij}  A^i A^j +  ( k \delta_{i}^j+ {\epsilon_{i}}^{j3}) A^i B_j + ( k \delta^{i}_j- {\epsilon^{i}}_{j3}) B_i A^j  + (   \delta^{ij}+\delta^{lk} \epsilon^i_{l3} \epsilon^j_{k3}) B_i B_j \bigr] . 
  \ee
The Lagrangian one-form is therefore:
\be
{\boldsymbol{\theta}}_{L}=(k\, {\cal \eta}_{IJ}+  {\cal H}_{IJ})\dot {\bf Q}^I  {\boldsymbol{\alpha}}^J   
\ee
 and  the equations of motion read as:
\be
{\sf L}_\Gamma\dot {\bf Q}^I( k\, {\cal \eta}_{IJ}+ \, {\cal H}_{IJ}) -\dot {\bf Q}^P \dot {\bf Q}^Q C_{IP}^K ( k\, {\cal \eta}_{QK}+ \,{\cal H}_{QK})  =0 \label{eomd}
\ee
where $C_{IP}^K$ are the structure constants of $\mathfrak{sl}(2,\C)$. The   matrix   $ k\, {\cal \eta}_{IJ}+ \, {\cal H}_{IJ}$ is non-singular, provided $k^2 \ne 1$, which will be assumed from now on. 
\subsection{Recovering the Standard Description}\label{standardlag}
 The standard  dynamics of the  isotropic rigid rotator is now shown to be recovered from the new Lagrangian.

To be definite, let us  fix a local decomposition for the elements of the double group $SL(2,\mathbb{C})$: $\gamma= \tilde{g}g$, with $g \in SU(2)$ and $\tilde{g} \in SB(2,\mathbb{C})$. 
  From eq. \eqref{newac}, one can see that $  L$ is invariant under left and  right action of the group $SU(2)$, 
 but only under left action of the group $SB(2,\mathbb{C})$, given by 
 \be 
 SB(2,\mathbb{C})_L: \gamma \rightarrow \tilde{h} \gamma=\tilde{h}\tilde{g}g, \quad \forall\tilde{h} \in SB(2,\mathbb{C}).
 \ee
In order to recover the usual description of the rotator,  the $SB(2,\C)_L$ invariance has to be promoted to a gauge symmetry.
One has then:
\be\label{sbgauge}
\gamma^{-1}d\gamma\rightarrow \gamma^{-1}D_{\tilde C}\gamma= (\gamma^{-1}\dot\gamma+\gamma^{-1}\tilde C\gamma) dt   
\ee
with 
\be
\tilde C= \tilde {C_i}(t)\tilde e^i \label{gaug}
\ee
the gauge connection. 
The following split can be performed:
\be
\gamma^{-1}\dot\gamma+\gamma^{-1}\tilde C\gamma=\gamma^{-1}\dot\gamma+ \tilde{C_i} \gamma^{-1} \tilde e^i \gamma= \mathcal{U}_i\tilde e^i + \mathcal{W}^i e_i   \nonumber \,\,\,.
\ee
Then, eq. \eqn{eupedown} implies:
\beqa
\mathcal{U}_i&=& 2{\rm Im}\Tr[(\gamma^{-1}\dot\gamma+\tilde{C_j} \gamma^{-1} \tilde e^j \gamma) e_i]= B_i+ \tilde{C_j} \;2{\rm Im}\Tr(\gamma^{-1} \tilde e^j \gamma  e_i) \label{Ui} \\
\mathcal{W}^i&=& 2{\rm Im}\Tr[(\gamma^{-1}\dot\gamma+\tilde{C_j}\gamma^{-1} \tilde e^j \gamma)  \tilde e^i]= A^i+ \tilde{C_j} \;2{\rm Im}\Tr(\gamma^{-1} \tilde e^j \gamma  \tilde e^i) \,\,\, .\label{Vi}
\eeqa
Let us explicitly compute the two terms in the r.h.s. of Eqs. \eqn{Ui}, \eqn{Vi} corresponding to the adjoint action of SL(2;C), in the chosen parametrization. After observing that the infinitesimal adjoint action of $g, \tilde g$ on $e_j, \tilde e^j$ is represented by the Lie brackets \eqn{su}, \eqn{sb}, \eqn{liemis}, one gets:
\be
\Tr(\gamma^{-1} \tilde e^j \gamma  e_i) = \Tr[(\tilde g^{-1} \tilde e^j \tilde g)(g  e_i g^{-1})] = \Tr (Ad_{\tilde g}\tilde e^j) (Ad_{g^{-1}} e_i )]= \Tr[( { {\rm a}(\tg)}^j_k\tilde e^k)({ {\rm h^{-1}}(g)}_{i}^ s e_s)]  \nonumber
\ee
so that,  from \eqn{psd} we have
\be
2{\rm Im}\Tr(\gamma^{-1} \tilde e^j \gamma e_i)= { {\rm a}^j}_k { {\rm h^{-1}}_i}^ s\delta_s^k  \nonumber
\ee
which yields:
\be
\mathcal{U}_i= B_i+ \tilde{C_j} { {\rm a}}^j_k ( { {\rm h^{-1}}})_{i }^k.  
\ee
Analogously, one can compute:
\beqa
\Tr(\gamma^{-1} \tilde e^j \gamma \tilde e^i)&=& \Tr[(\tilde g^{-1} \tilde e^j \tilde g)(g \tilde e^i g^{-1})] = \Tr (Ad_{\tilde g}\tilde e^j) (Ad_{g^{-1}} \tilde e^i )]\nn\\
&=& \Tr[( { {\rm a}(\tg)}^j_k\tilde e^k)({ {\rm b^{-1}}(g)}_s^i \tilde e^s+{ {\rm d^{-1}}(g)}^{i s} e_s)]  \nonumber
\eeqa
and, from  \eqn{psd}:
\be
2{\rm Im}\Tr(\gamma^{-1} \tilde e^j \gamma \tilde e^i)= { {\rm a}(\tg)}^j_k { {\rm d^{-1}}(g)}^{i s}\delta^k_s  \nonumber
\ee
that is 
\be
\mathcal{W}^i= A^i+ \tilde{C_j} { {\rm a}}^j_k { {\rm d}}^{i k}. \label{calU}
\ee
After replacing the Lagrangian in \eqn{newac} with the gauged Lagrangian
\be
L_{\tilde C}=\frac{1}{2}\bigl[ k \Braket{\gamma^{-1}\mathrm{D}\gamma\stackrel{\wedge}{,}* \gamma^{-1}\mathrm{D}\gamma} +((\gamma^{-1}\mathrm{D}\gamma \stackrel{\wedge}{,} * \gamma^{-1}\mathrm{D} \,\,\,,
\gamma)) \bigr]   
\ee
one gets:
\be
L_{\tilde C}=\frac{1}{2}  (k \,{\cal \eta}_{IJ} + {\cal H}_{IJ})\dot {\bf {\widehat Q}}^I  \dot {\bf \widehat Q}^J 
\ee
with 
\be
\dot {\bf \widehat Q}^I= ({\mathcal{W}^i, \mathcal{U}_i})
\ee
namely
\be
L_{\tilde C}=\frac{1}{2} \bigl[  \delta_{ij}  \mathcal{W}^i \mathcal{W}^j + 2 (k  \delta_{i}^j+ \epsilon_{i}^{j3}) \mathcal{W}^i \mathcal{U}_j +(   \delta^{ij}+\delta^{lk} \epsilon^i_{l3} \epsilon^j_{k3}) \mathcal{U}_i \mathcal{U}_j \bigr] \,\, .
  \ee
 Let us introduce now the combination:
 \be\label{tildeV}
 \widehat{\mathcal{ W}}^i= \mathcal{ W}^i+  (k \delta^{is} -\epsilon^{is}_3) \mathcal{U}_s \,\,\, ,
 \ee
allowing to rewrite the Lagrangian $L_{\tilde C}$ as follows:
 \be
L_{\tilde C}=\frac{1}{2} \bigl[  \delta_{ij}  \widehat{\mathcal{W}}^i \widehat{\mathcal{W}}^j + (1-k^2)\delta^{ij}{\mathcal U}_i{\mathcal U}_j\bigr] \,\,\,.  
\ee
This can be used for writing the partition function of the system under analysis as:
\begin{equation}
Z=\int \mathcal{D}g \mathcal{D}\tilde{g}\mathcal{D}{\tilde C} e^{-S_{\tilde C}}
\label{partitt}
\end{equation}
and integrate over the gauge potential.  Therefore,  the integration with respect to $\tilde{C_i}$ can be traded for the integration with respect to ${\mathcal U}_i$. 
The functional integral \eqref{partitt} can be performed by changing the integration variable. Therefore, by inverting the relation \eqref{calU}, one can calculate $ \det\biggl(\frac{\delta \tilde{C_i}}{\delta {\mathcal U}_j}\biggr)$ and see that it is a constant, because the matrices  involved in the definition of ${\mathcal U}_i$ are all invertible. Consequently, the functional integral in the partition function becomes:
\begin{equation}
Z=\int \mathcal{D}g \mathcal{D}\tilde{g} e^{-\frac{1}{2}\int_{\mathbb{R}}\mathrm{d}t ( \delta_{ij}\widehat{\mathcal{W}}^i \widehat{\mathcal{W}}^j)} \int \mathcal{D}{\mathcal U} e^{-\frac{1}{2}\int_{\mathbb{R}}\mathrm{d}t(1-k^2)\delta^{ij}{\mathcal U}_i {\mathcal U}_j},  
\label{patint}
\end{equation}
where
\begin{equation}
\int \mathcal{D}{\mathcal U} e^{-\frac{1}{2}(1-k^2) \int_{\mathbb{R}}\delta^{ij}{\mathcal U}_i {\mathcal U}_j}=(2\pi)^{\frac{3}{2}}(\det(\delta^{ij}))^{-\frac{1}{2}}.
\label{fint}
\end{equation}

It is worth noticing that, in \eqn{tildeV}, the tensor $T^{ij}=k \delta^{ij}- \epsilon^{ij3}$ defines, for $k\neq 0,$ a constant invertible map $T: \mathfrak{sb}(2,\mathbb{C}) \rightarrow \mathfrak{su}(2),$ so one can introduce the endomorphism $E$ of $\mathfrak{d}= \mathfrak{su}(2) \oplus \mathfrak{sb}(2,\mathbb{C})$ which preserves the splitting, defined by the constant matrix:
\begin{equation} 
E^I_J=
\begin{pmatrix}
\delta^i_j & T^{ij} \\
-(T^{-1})_{ij} &  \delta_i^j
\end{pmatrix}
\label{endom}
\end{equation}
This acts on any element of $\mathfrak{d}$ in the following way:
\begin{equation} 
\begin{pmatrix}
\delta^i_j & T^{ij} \\
-(T^{-1})_{ij} &  \delta_i^j
\end{pmatrix}
\begin{pmatrix}
\mathcal{W}^j \\
\mathcal{U}_j 
\end{pmatrix}
=
\begin{pmatrix}
\widehat{\mathcal{W}}^i \\
\widehat{\mathcal{U}}_i 
\end{pmatrix}    \nonumber
\end{equation}
where $\widehat{\mathcal{W}}^i$ is given by \eqn{tildeV} and $\widehat{\mathcal{U}}_i=\mathcal{U}_i - (T^{-1})_{ij} \mathcal{W}^j.$  We can write down the left invariant forms $$g'^{-1} \mathrm{d} g' = \widehat{\mathcal{W}}^i e_i \mathrm{d}t$$ and 
$$\tilde{g}'^{-1} \mathrm{d} \tilde{g}' = \widehat{\mathcal{U}}_i \tilde{e}^i \mathrm{d}t.$$

The constant endomorphism \eqn{endom} induces a map $\exp(E): SL(2,\mathbb{C}) \rightarrow SL(2,\mathbb{C})$  such that $\gamma= \tilde{g}g \rightarrow \gamma' = \tilde{g}' g'.$ Then, one can see that the path integral measure can be transformed giving $\mathcal{D}g \mathcal{D}\tilde{g} = \mathcal{D}g' \mathcal{D}\tilde{g}'$ up to a constant factor, i.e. the determinant of the constant map $\exp(E)$. 

Thus the path integral \eqn{patint} can be written, up to constant factors, as:
\be
Z=\int \mathcal{D}\tilde{g}' \int \mathcal{D}g' e^{-\frac{1}{2} \int_{\mathbb{R}} \Tr[g'^{-1} \mathrm{d} g' \wedge *g'^{-1} \mathrm{d} g' }] 
\ee
where the path integral over $\tilde{g}'$ gives a constant and the other integral is exactly the partition function of the action of the IRR defined up to a constant factor.

\subsection{Recovering the Dual Model}\label{recdu}
The dual model described by the action functional \eqn{dualag} can be recovered along the same lines as in the previous section. We consider the parent action \eqn{newac}, with the same parametrization as before, namely $\gamma=\tilde g g$,  and explore the global invariance under right $SU(2)$ action 
\be 
 SU(2)_R: \gamma \rightarrow \gamma h=\tilde{g}g h , \quad \forall {h} \in SU(2).
 \ee
Hence,   in  complete analogy with eq. \eqn{sbgauge},   we gauge this symmetry, by introducing the $\mathfrak{su}(2)$-valued  connection one-form $ C= C^i (t) e_i$ so that
\be\label{sugauge}
\gamma^{-1}d\gamma\rightarrow \gamma^{-1}D\gamma= (\gamma^{-1}\dot\gamma+\gamma^{-1} C\gamma) dt   
\ee
Notice that in this case we could gauge the left $SU(2)$ action, in which case it would be  convenient  to use the other parametrization of $\gamma$ as $\gamma =k  \tilde k , ~k\in SU(2), \tilde k\in SB(2,\C)$. 

From \eqn{sugauge} we have:
\be
\gamma^{-1}D\gamma=\tilde {\mathcal U}_i {\tilde e}^i + \tilde {\mathcal W}^i e_i 
\ee
with 
\beqa
\tilde{\mathcal U}_i&=& 2{\rm Im}\Tr[(\gamma^{-1}\dot\gamma+ C^j \gamma^{-1} e_j \gamma)e_i]= B_i+ C^j \;2{\rm Im}\Tr(\gamma^{-1} { e}_j \gamma e_i) \label{tUi} \\
\tilde{\mathcal W}^i&=& 2{\rm Im}\Tr[(\gamma^{-1}\dot\gamma+C^j \gamma^{-1} e_j \gamma)  \tilde e^i]= A^i+  C^j \;2{\rm Im}\Tr(\gamma^{-1} e_j \gamma  \tilde e^i) \,\,\, .\label{tVi}
\eeqa
By using the adjoint action of $SL(2,\C)$ on $e_i, \tilde e^i$, we obtain:
\be
\Tr(\gamma^{-1} e_j \gamma  e_i) = \Tr[(\tilde g^{-1} e_j \tilde g)(g  e_i g^{-1})] = \Tr (Ad_{\tilde g}e_j) (Ad_{g^{-1}} e_i )]= \Tr[\left( { {\rm l}(\tg)}_j^k  e_k + {\rm m }_{jk}(\tilde g)\tilde e^k\right)\left({ {\rm h^{-1}}(g)}_{i}^ s e_s\right)]  
\ee
so that,  from \eqn{psd} one gets:
\be
2{\rm Im}\Tr(\gamma^{-1} e_j \gamma  e_i) =  {\rm m}_{jk} {( {\rm h^{-1}})_i}^ k  \nonumber
\ee
which yields:
\be
\tilde{\mathcal{U}}_i= B_i+  C^j {\rm m}_{jk} { ({\rm h^{-1}})_i}^ k.  \label{caltU}
\ee
Analogously, one can compute:
\beqa
\Tr(\gamma^{-1}e_j \gamma \tilde e^i)&=& \Tr[(\tilde g^{-1} e_j \tilde g)(g \tilde e^i g^{-1})] = \Tr (Ad_{\tilde g}e_j) (Ad_{g^{-1}} \tilde e^i )]\nn\\
&=& \Tr[( { {\rm l}(\tg)}_j^ke_k+{\rm m }_{jk}(\tilde g) \tilde e^k )({ {\rm b^{-1}}(g)}_s^i \tilde e^s+{ {\rm d^{-1}}(g)}^{i s} e_s)]  \nonumber
\eeqa
and, from  \eqn{psd}
\be
2{\rm Im}\Tr(\gamma^{-1}e_j \gamma \tilde e^i)= {\rm l}_j^k ({\rm b^{-1}})_k^i +  { {\rm m}}_j^k ( {\rm d^{-1}})^{i k}   \nonumber
\ee
that is 
\be
\tilde{ \mathcal{W}}^i= A^i+ \tilde C^j\left ( {\rm l}_j^k ({\rm b^{-1}})_k^i +  { {\rm m}}_j^k ( {\rm d^{-1}})^{i k}  \right) \label{caltW}
\ee
The gauged Lagrangian reads then
\be
{ L}_{ C}= \frac{1}{2}\left(k {\cal \eta}_{IJ}+ {\cal H}_{IJ}\right){\dot{\tilde {\mathbf  Q}}^I} {\dot{\tilde {\mathbf  Q}}^J}
\ee
with $\dot{\tilde {\mathbf { Q}}}^I \equiv \left( \tilde{\mathcal{W}}^i, \tilde{\mathcal{U}}_i\right) $
so that
\beqa
{ L}_{  C}&=& \frac{1}{2}\left[\delta_{ij} \tilde{\mathcal{W}}^i \tilde{\mathcal{W}}^j + 2 (k \delta_i^j +{ \epsilon_i}^{j3}) \tilde{\mathcal{W}}^i \tilde{\mathcal{U}}_j + 
(\delta^{ij} + \delta^{lk}{ \epsilon^i}_{l3}{\epsilon^j}_{k3}) \tilde{\mathcal{U}}_i \tilde{\mathcal{U}}_j \right]\nn\\
&=& \frac{1}{2}\left[\delta_{ij} \tilde{\mathcal{W}}^i \tilde{\mathcal{W}}^j + 2 (k \delta_i^j +{ \epsilon_i}^{j3}) \tilde{\mathcal{W}}^i \tilde{\mathcal{U}}_j + 
h^{ij} \tilde{\mathcal{U}}_i \tilde{\mathcal{U}}_j \right].
\eeqa
We can proceed as in previous section and introduce  
 \be\label{tildeUdu}
 \breve{\mathcal{ U}}_i= \tilde{ \mathcal{ U}}_i+  \tilde{ \mathcal{W}}^s {T_s}^l (h^{-1})_{il} \,\,\, ,
 \ee
 with 
 \be
 {T_s}^l = k\delta_s^l+{\epsilon_s}^{l3}
 \ee
  and the inverse metric 
 \be
 (h^{-1})_{il}= \delta_{il}-\frac{1}{2}\delta_{pq}{\epsilon_i}^{p3} {\epsilon_l}^{q3}
 \ee
allowing to rewrite the Lagrangian ${\widehat L}_C$ as follows:
 \be
{L}_{ C}=\frac{1}{2} \left[  \left(\delta_{ij}- {T_i}^k  {T_j}^l(h^{-1})_{kl}  \right) \tilde{\mathcal{W}}^i \tilde{\mathcal{W}}^j +h^{ij}
 \breve{\mathcal{U}}_i \breve{\mathcal{U}}_j \right]\,\,\,.  
\ee
Thus we can write  the partition function of the system under analysis as:
\begin{equation}
Z=\int \mathcal{D}g \mathcal{D}\tilde{g}\mathcal{D}{ C} e^{-S_{ C}} 
\label{partittdu}
\end{equation}
and integrate over the gauge potential. 

 Let us stress that the difference with respect to the previous case is that now the gauge connection is an $SU(2)$ one, therefore allowing to trade the integration over ${C}^i$ for the integration over $\tilde{\mathcal{W}}^i $. 

By repeating exactly the same steps as in Sec.  \ref{standardlag}, one arrives at:
\begin{equation}
Z=\int \mathcal{D}g \mathcal{D}\tilde{g} e^{-\frac{1}{2}\int_{\mathbb{R}}\mathrm{d}t h^{ij}\breve{\mathcal U}_i \breve{\mathcal U}_j}  \int \mathcal{D}\breve{\mathcal{W}} e^{-\frac{1}{2}\int_{\mathbb{R}}\mathrm{d}t  \left(\delta_{ij}- {T_i}^k  {T_j}^l(h^{-1})_{kl}  \right) \breve{\mathcal{W}}^i \breve{\mathcal{W}}^j},  
\label{patintdu}
\end{equation}
and
\begin{equation}
\int \mathcal{D}\breve{\mathcal{W}} e^{-\frac{1}{2}\int_{\mathbb{R}}\mathrm{d}t  \left(\delta_{ij}- {T_i}^k  {T_j}^l(h^{-1})_{kl}  \right) \breve{\mathcal{W}}^i \breve{\mathcal{W}}^j}=(2\pi)^{\frac{3}{2}}\left(\det (\delta_{ij}- {T_i}^k  {T_j}^l(h^{-1})_{kl} ) \right) ^{-\frac{1}{2}}.
\label{fintdu}
\end{equation}
It is worth noticing that, as in \eqn{tildeV}, also in \eqn{tildeUdu} the tensor $\breve{T}_{ij}=(h^{-1})_{il} T^l_j$ defines a constant invertible map $\breve{T}:\mathfrak{su}(2) \rightarrow \mathfrak{sb}(2,\mathbb{C}),$ so that we can use the  split-preserving endomorphism $\breve{E}$   of $\mathfrak{d}= \mathfrak{su}(2) \oplus \mathfrak{sb}(2,\mathbb{C})$, defined below, to get:
\begin{equation} 
\breve{E}\begin{pmatrix}
\tilde{\mathcal{W}}^j \\
\tilde{\mathcal{U}}_j 
\end{pmatrix}=\begin{pmatrix}
\delta^i_j & -(\breve{T}^{-1})^{ij} \\
\breve{T}_{ij} &  \delta_i^j
\end{pmatrix}
\begin{pmatrix}
\tilde{\mathcal{W}}^j \\
\tilde{\mathcal{U}}_j 
\end{pmatrix}
=
\begin{pmatrix}
\breve{\mathcal{W}}^i \\
\breve{\mathcal{U}}_i 
\end{pmatrix}    
\end{equation}
where $\breve{\mathcal{U}}_i$ is given by \eqn{tildeUdu} and $\breve{\mathcal{W}}^i=\tilde{\mathcal{W}}^i - (\breve{T}^{-1})^{ij} \tilde{\mathcal{U}}_j.$  
The constant endomorphism $\breve{E}$ induces a map $\exp(\breve{E}): SL(2,\mathbb{C}) \rightarrow SL(2,\mathbb{C})$  which preserves the chosen parametrization, namely,  
$\exp(\breve{E}): \gamma= \tilde{g}g \rightarrow \gamma' = \tilde{g}' g'.$ Then, one can see that the path integral measure can be transformed giving $\mathcal{D}g \mathcal{D}\tilde{g} = \mathcal{D}g' \mathcal{D}\tilde{g}'$ up to a constant factor, i.e. the determinant of the constant map $\exp(\breve{E})$.
Finally, by introducing the left invariant forms $$g'^{-1} \mathrm{d} g' = \breve{\mathcal{W}}^i e_i \mathrm{d}t$$ and 
$$\tilde{g}'^{-1} \mathrm{d} \tilde{g}' = \breve{\mathcal{U}}_i \tilde{e}^i \mathrm{d}t$$
 the path integral \eqn{patintdu} can be written, up to constant factors, as:
\be
Z=\int \mathcal{D}{g}' \int \mathcal{D}\tilde{g}' e^{-\frac{1}{2}\int_{\mathbb{R}}{\mathcal Tr} [{\tilde g}'^{-1} \mathrm{d} {\tilde g}' \wedge *\tilde{g}'^{-1} \mathrm{d}\tilde{ g}' } ]
\ee
where the path integral over ${g}'$ gives a constant and the other integral is exactly the partition function  of the dual model defined up to a constant factor.

\subsection{The Hamiltonian Formalism}\label{hamform} 
In the doubled description introduced above, the left generalized momenta are represented by:
\be
{\bf P}_I = \frac{\del \widehat L}{\del \dot {\bf Q}^I}= ( {\cal \eta}_{IJ}+ k\,{\cal H}_{IJ})\dot {\bf Q}^J  \label{genP}
\ee
The Hamiltonian reads then as:
\be
\widehat{H}= ({\bf P}_I \dot{\bf Q}^I - \widehat L)_{{\bf P}}= \frac{1}{2} [( {\cal \eta}+ k\, {\cal H})^{-1}]^{IJ} {\bf P}_I{\bf P}_J   \nonumber
\ee
with 
\be
[( {\cal \eta} + k\, {\cal H})^{-1}]^{IJ}= \frac{1}{2} (1-k^2)^{-1} 
\begin{pmatrix}
\delta^{ij}+ \epsilon^i_{l3} \epsilon^j_{k3}\delta^{lk}& -{\epsilon^i}_{j3}-k \delta^i_j
\\
{\epsilon_i}^{j3}-k \delta_i^j& \delta_{ij} \,\,\, 
\end{pmatrix}  \,\,.    \nonumber
\ee
From  \eqn{genP} one can explicitly write the generalized momenta ${\bf P}_I$ in terms of the components of $\dot{\bf Q}^I\equiv(A^i, B_j)$, finding:
\be
{\bf P}_I \equiv ( I_i, \tI^i)=\left(\delta_{ij} A^j+(k\delta_i^j+ \epsilon_i^{j3})B_j, (k \delta^i_j-\epsilon^i_{j3})A^j+[\delta^{ij}+\delta^{lk}\epsilon^i_{l3}\epsilon^j_{k3}]B_j\right).  \nonumber
\ee
In terms of the components $I_i, \tI^j$, it turns out that:
\beqa
\widehat{H} &=&\frac{1}{2}(1-k^2)^{-1}\left( \delta^{ij} I_i I_j + \delta_{ij} \tI^i \tI^j + \epsilon^i_{l3} \epsilon^j_{k3}\delta^{lk} I_i I_j -2 k   \delta^i_j I_i \tI^j + 2\epsilon_i^{j3} \tI^iI_j \right) \nonumber \\
&=& 
\frac{1}{2}(1-k^2)^{-1} \left ( (1-k^2) \delta^{ij} I_i I_j + \delta_{ij}(\tI^i -I_s(k \delta^{si}+ {\epsilon^{si}}_3))
(\tI^j -I_r(k  \delta^{rj}+ {\epsilon^{rj}}_3))\right)\nn
\eeqa
which can be rewritten as
\be
{\widehat{H}} =\frac{1}{2}(1-k ^2)^{-1}\left((1-k ^2) \delta^{ij} I_i I_j + \delta_{ij}\tilde{\cal I}^i \tilde{\cal I}^j\right)  \nonumber
\ee
after having defined
\be
\tilde{\cal I}^i \equiv \tI^i -I_s(k  \delta^{si}+ {\epsilon^{si}}_3)= \delta^{ij} (1-k^2) B_j.  \nonumber
\ee
In order to obtain the Hamilton equations for the generalized model on the Drinfel'd double,  one can proceed as in the previous section with the determination of Poisson brackets from the first-order action functional:
\be
{\widehat{\mathcal{S}}}= \int \langle {\bf P} |  \gamma^{-1}d \gamma\rangle  - \int \widehat{H} dt \equiv \int  {\boldsymbol{\theta}} -\int \widehat{H} dt  \nonumber
 \ee
with 
\beqa
{\bf P}&=& i\; {\bf P}_I {e^I}^*= i\;(I_i{e^i}^* + \tI_i \te_i^*) \nonumber \\
\gamma^{-1}d\gamma&=& i\,  {\boldsymbol{\alpha}}^J e_J= (\alpha^k e_k + \beta_k \te^k)  \nonumber  \,\, .
\eeqa
We stress once again that ${\bf P_I}$, $ {\boldsymbol{\alpha}}^J$ are respectively generalized momenta and basis one-forms on the doubled configuration space $SL(2,\C)$. The symplectic form on $T^*SL(2,\C)\simeq SL(2,\C)\times \mathfrak{sl}(2,\C)^*$ is therefore:
\beqa
{\boldsymbol{\omega}}= d {\boldsymbol{\theta}}&=& dI_i\wedge \alpha^i + d\tI^i\wedge  \beta_i +\frac{1}{2}\tI^l\left(\alpha^j\wedge\beta_k {\epsilon^k}_{jl}- \beta_j\wedge \alpha^k {\epsilon^j}_{kl}- \beta_j\wedge\beta_k {f^{jk}}_l\right)\nn\\
&+&  \frac{1}{2}I_l\left(-\alpha^j\wedge\alpha^k {\epsilon^l}_{jk}+ \alpha^j\wedge \beta_k {f^{lk}}_{j}- \beta_j\wedge\alpha^k {f^{lj}}_k\right)  \nonumber
\eeqa
which  yields for the generalized momenta the Poisson brackets:
\beqa \label{remark}
\{I_i, I_j\}&=& {\epsilon_{ij}}^k I_k \\
\{\tI^i, \tI^j\}&=& {f^{ij}}_k \tI^k\\
\{I_i, \tI^j\}&=& {\epsilon^j}_{il} \tI^l- I_l {f^{lj}}_i  \;\;\;\;\{\tI^i, I_j\}= -{\epsilon^i}_{jl} \tI^l+ I_l {f^{li}}_j \label{remark3}
\eeqa
while the Poisson brackets between momenta and configuration space variables $g,\tg$ are unchanged with respect to $T^*SU(2), T^*SB(2,\C)$. We shall come back to the Poisson algebra \eqn{remark} in the next subsection.

In order to derive Hamilton equations, it is sufficient to write in compact form:
\be
\{{\bf P}_I, {\bf P}_J\}= C_{IJ}^K {\bf P}_K   \nonumber
\ee
with $C_{IJ}^K$ the structure constants specified above. We have then: 
\be
\frac{d}{dt} {\bf P}_I= \{ {\bf P}_I, \widehat H\}=  [( {\cal \eta}+ k\, {\cal H})^{-1}]^{JK} \{ {\bf P}_I, {\bf P}_J\} {\bf P}_K= [( {\cal \eta}+ k\, {\cal H})^{-1}]^{JK} C_{IJ}^L {\bf P}_L{\bf P}_K   \nonumber
\ee
which is not zero, consistently with \eqn{eomd}.
 
\subsection{The Poisson Algebra}\label{canform}
The generalized formulation of the isotropic rotator is completed by discussing  the Poisson brackets on the double group $SL(2,\C)$, which correctly generalize those on the cotangent bundle stated in eq.s  \eqn{pp}-\eqn{ij} as well as in Eqs. \eqn{ppdu}-\eqn{ijdu}. These have been introduced long time ago in \cite{semenov, alex} in the form
\be
\{\gamma_1,\gamma_2\}= -\gamma_1\gamma_2 r^* -r \gamma_1\gamma_2 \label{gammagamma2}
\ee
where $\gamma_1= \gamma\otimes 1, \gamma_2= 1\otimes \gamma_2$ while $r \in \mathfrak{d} \otimes \mathfrak {d}$ is the classical Yang-Baxter matrix:
 \be
r =  e^i\otimes e_i \label{rmatrix}
\ee 
satisfying the modified Yang-Baxter equation 
\be
[r_{12},r_{13}+r_{23}]+ [r_{13},r_{23}]= h   \nonumber
\ee
with  $r_{12 }=e^i\otimes e_i  \otimes \mathds{1}$, $r_{13}=e^i\otimes\mathds{1}\otimes  e_i $, $r_{23}= \mathds{1\otimes e^i\otimes e_i }$,  and  $h\in \mathfrak{d}\otimes \mathfrak{d}\otimes\mathfrak{d}$ and adjoint invariant element in the enveloping algebra.  The matrix 
\be
r^*= - e_i\otimes e^i \label{rmatrix2}
\ee
is also solution of the Yang-Baxter equation. The group $D$ equipped with the Poisson bracket \eqn{gammagamma2} is also called the Heisenberg double \cite{semenov,alex}. On writing $\gamma$ as $\gamma= \tilde g g$ it can be shown that \eqn{gammagamma2} is compatible with the following choice 
\begin{align} 
 \{g_1,g_2\}  &= [r^*,g_1g_2], 
\label{pbm1}\\
\{{\tilde g}_1,g_2\} &=- {\tilde g}_1r  g_2 \label{finalpoi}\\
 \{\tilde g_1,\tilde g_2\}  &=-[r,\tilde g_1\tilde g_2],
\label{pbm2} 
\end{align}
with $g_1=g\otimes \mathds{1}$, $g_2=\mathds{1}\otimes g$, $\tilde g_1={\tilde g}\otimes \mathds{1}$ and ${\tilde g}_2= \mathds{1} \otimes {\tilde g}$.  eq.s \eqn{pbm2} \eqn{pbm1} are the so-called Sklyanin brackets \cite{skly}. We also have $\{ { g}_1,{\tilde g}_2\} =- {\tilde g}_2 r^*   g_1$. 

Let us verify that we actually recover eq.s \eqn{pp}-\eqn{ij}. In order to obtain the PB on the fibers of the cotangent bundle $T^*SU(2)$, the matrix $r$  is rescaled by a real parameter  $\lambda$ and  the elements of $G^*$ are made dependent on the same parameter. By expanding up to the first order, one gets: 
\begin{equation} 
\tilde g(\lambda)=e^{i\lambda I_i e^i}   = 1+i\lambda I_i e^i + \mathcal{O}(\lambda^2).
\label{expansion1}
\end{equation}
Substituting this in \eqref{pbm2} yields, for the left-hand side:
\begin{equation}
\{\tilde g_1,\tilde g_2\}=\{\tilde g \otimes \mathds{1}, \mathds{1}\otimes \tilde g\}\simeq -\lambda^2 e^i\otimes e^j \{I_i,I_j\} + \mathcal{O}(\lambda^3),   \nonumber
\end{equation}
and  for the right-hand side:
\begin{equation} \label{calc1}
\begin{split}
[r,\tilde g_1 \tilde g_2]\simeq & -\lambda \left([e^i, i \lambda  I_j e_j]  \otimes e_i +  e^i \otimes [e_i, i \lambda I_j e^j]\right) + \mathcal{O}(\lambda^3) \\
= & -i \lambda^2 I_j  \bigl([e^i,e_j]\otimes e_i + e^i \otimes [e_i,e^j]\bigr)+ \mathcal{O}(\lambda^3) \\
= &  \lambda^2 I_j (f^{ij}_r e^r \otimes e_i -  \epsilon^j_{ir}e^i\otimes e^r-f^{rj}_i e^i\otimes e_r)+ \mathcal{O}(\lambda^3) \\
= & - \lambda^2 I_k  \epsilon^k_{\,ij}e_i\otimes e_j+ \mathcal{O}(\lambda^3).  
\end{split}
\end{equation}
By equating the two sides, in the limit $\lambda \rightarrow 0$,  one obtains the Poisson bracket:
\begin{equation}
\{I_i,I_j\}=\epsilon^k_{\,ij}I_k.  \label{first}
\end{equation}
Let us consider the second Poisson bracket, eq. \eqn{finalpoi}. In order to compute its l.h.s. we use for $g$ the parametrization $g= y^0 \sigma_0 + i y^i \sigma_i$. We have, up to the first order in $\lambda$:
\be
\{\tilde g_1,g_2\}=2 i\lambda\left( \{I_i, y^0\} e^i\otimes e_0+i \{I_i, y^j\} e^i\otimes e_j\right)+ O(\lambda^2)\label{lhs2}
\ee
while for the r.h.s.
\beqa
-\tilde g_1 r g_2&\simeq& -2 \left((\mathds{1} + i\lambda I_i e^i )\otimes \mathds{1}\right)(\lambda e^k\otimes e_k)\left(\mathds{1}\otimes (y^0 e_0 + i y^j e_j)\right) \nn\\
&=& -2\lambda e^k\otimes e_k \left(\mathds{1}\otimes (y^0 e_0 + i y^j e_j)\right)+ O(\lambda^2)\nn\\
&=& -2 \lambda( \frac{1}{2}y^0 e^k\otimes e_k+ i y^j e^k\otimes e_k e_j)+ O(\lambda^2)\nn\\
&=& - \lambda(  y^0 e^k\otimes e_k+ i y^j e^k\otimes (\delta_{kj}e_0+ i \epsilon_{kj}^i e_i)\label{rhs2}
\eeqa
After equating  \eqn{lhs2} with \eqn{rhs2},
one finally gets at order $\lambda$:
\beqa\label{second}
\{I_i, y^0\}&=& - y^j \delta_{ij}  \nonumber \\
\{I_i, y^j\}&=&  y^0 \delta_i^j - y^k \epsilon_{ki}^j   \nonumber
\eeqa
where the first one is compatible with the second one, by using $(y^0)^2= 1- \sum_k y^k y^k$. 
Finally, let us consider \eqref{pbm1}. The l.h.s. yields:
\be
\{{g}_1, {g}_2\}= \{y^0, y^j\} i (\sigma_0\otimes \sigma_j-\sigma_j\otimes \sigma_0) -  \{y^i, y^j\}  \sigma_i\otimes \sigma_j
\label{lhsgg}
\ee
which does not depend on $\lambda$. The r.h.s. instead reads as:
\be
[r^*, g_1g_2]= -\lambda[e_k\otimes e^k, g\otimes g]+O(\lambda^2)\label{rhsgg}
\ee
which is at least first order in $\lambda$. Therefore, by comparing \eqn{lhsgg} with \eqn{rhsgg}, one obtains:
\be
\{y^0, y^j\} = \{y^i, y^j\} = 0 + O(\lambda) \,\,\, . \label{third}
\ee
Thus, Eqs. \eqn{first}, \eqn{second}, \eqn{third} reproduce correctly the canonical Poisson brackets on the cotangent bundle in Eqs. \eqn{pp}-\eqn{ij}. 

In order to underline the symmetric role played by the group $SU(2)$ and its dual, one can perform a slightly different analysis by considering $r^*$ as an  independent solution of the Yang-Baxter equation  
\be
\rho= -\mu e_k\otimes e^k \label{altrmat}
\ee
and expanding $g\in SU(2)$ as a function of the parameter $\mu$:
\be
g= \mathds{1} + i \mu \tI^i e_i + O(\mu^2)   \label{gexp} \,\,\, .
\ee
By repeating the same analysis as above, it is straightforward to prove that the Poisson structure induced by $\rho$ is the one that correctly reproduces the canonical Poisson brackets on the cotangent bundle of $G^*=SB(2,\C)$ derived in Eqs. \eqn{ppdu}-\eqn{ijdu}. Indeed, by substituting \eqn{gexp} in the LHS of \eqn{pbm1}  one finds:
\begin{equation}
\{ g_1, g_2\}\simeq -\mu^2 e_i\otimes e_j \{\tI^i,\tI^j\} + \mathcal{O}(\mu^3),   \nonumber
\end{equation}
and  for the right-hand side:
\begin{equation}\label{calcdu}
\begin{split}
[\rho, g_1  g_2]\simeq & -\mu \left([e_i, i \mu  \tI^j e_j]  \otimes e^i +  e_i \otimes [e^i, i \mu \tI^j e_j]\right) + \mathcal{O}(\mu^3) \\
= & -i \mu^2 \tI^j  \bigl([e_i,e_j]\otimes e^i + e_i \otimes [e^i,e_j]\bigr)+ \mathcal{O}(\mu^3) \\
= &  \mu^2 \tI^j (\epsilon_{ij}^r e_r \otimes e^i + { f^{ri}}_j e_i\otimes e_r+{\epsilon^i}_{jr} e_i\otimes e^r)+ \mathcal{O}(\mu^3) \\
= &  \mu^2 \tI^k  {f^{\,ij}}_k e_i\otimes e_j+ \mathcal{O}(\mu^3).  
\end{split}
\end{equation}
By equating the two sides, in the limit $\mu \rightarrow 0$,  one obtains the Poisson bracket:
\begin{equation}
\{\tI^i,\tI^j\}=f_k^{\,ij}\tI^k.  \label{scnd}
\end{equation}

Last but not least, it is possible to consider a different Poisson structure on the double, given by \cite{semenov} : 
\be
\{\gamma_1,\gamma_2 \}= \frac{\lambda}{2}\left[\gamma_1(r^*-r) \gamma_2 - \gamma_2(r^*-r)\gamma_1\right].\label{gammargamma}
\ee
This is the one that correctly dualizes  the bialgebra structure on $\mathfrak{d}$ when evaluated at the identity of the group $D$. To this, let us expand $\gamma\in D$ as $\gamma= \mathds{1}+ i\lambda I_i \tilde e^i+ i\lambda \tI^i e_i$  and rescale  $r, r^*$  by the same parameter $\lambda$. 
It is straightforward to obtain, on the l.h.s. of eq. \eqn{gammargamma},
\be
\{\gamma_1,\gamma_2 \}= -\lambda^2 \left(\{I_i, I_j\} \tilde e^i\otimes \tilde e^j + \{\tI^i, \tI^j\}  e_i\otimes  e_j  +\{I_i, \tI^j\} (\tilde e^i\otimes  e_j- e_j \otimes \tilde e^i)   \right)  \nonumber
\ee
while, on the r.h.s. of the same equation:
\be
-\lambda^2 \left(I_s \epsilon^s_{\,ij} \tilde e^i\otimes  \tilde e^j+ \tI^s f_s^{\,ij}  e_i\otimes   e_j+ I_s f_i^{\,sj}(\tilde e^i\otimes e_j-e_j\otimes \tilde e^i ) + \tI^s \epsilon^j_{\,si}(\tilde e^r\otimes  e_j -  e_j\otimes \tilde e^i)\right) \,\,.  \nonumber
\ee
By equating the two results one obtains: 
\beqa
\{I_i, I_j\}&=& {\epsilon_{ij}}^k I_k  \nonumber \\
\{\tI^i, \tI^j\}&=& {f^{ij}}_k \tI^k  \nonumber \\
\{I_i, \tI^j\}&=&  - {f_i}^{ jk}I_k - \tI^k {\epsilon _{ki}}^{ j}  \nonumber
\eeqa
which is nothing but the Poisson bracket induced by the Lie algebra structure of the double \eqref{liemis}. 

By using the  compact notation $I= i I_i {e^i}^*, \tI = i \tI_i {\te_i}^*$, one can rewrite the Poisson algebra as follows: 
\be
\{I +\tI, J+\tilde{J}\}= \{I,J\} -\{J, \tI\}+ \{I,\tilde{J}\} + \{\tI,\tilde{J}\}. \label{cb}
\ee
This is a very interesting structure, which represents a Poisson realization of the C-bracket for the generalized bundle $T\oplus T^*$ over $SU(2)$, once one considers the isomorphisms 
\be
TSL(2,\C)\simeq SL(2,\C)\times \mathfrak{sl}(2,\C)  \nonumber
\ee
with the fiber:
\be
 \mathfrak{sl}(2,\C)\simeq \mathfrak{su}(2)\oplus \mathfrak{sb}(2,\C)\simeq TSU(2)\oplus T^*SU(2).  \nonumber
 \ee
That is, we recognize $I= i I_i {e^i}^*, J= i J_i {e^i}^*$ as one-forms, with ${e^i}^*$ being a basis over $T^*$ and  $\tI = \tI^i {\te_i}^*, \tilde J= \tilde J^i {\te_i}^*$ as  vector-fields, with ${\te_i}^*$ a basis over $T$. Namely, the couple $(I_i, \tI^i)$ identifies the fiber coordinate of the generalized bundle $T\oplus T^*$ of $SU(2)$.

In order to complete the analysis, let us look at the Lie algebra of Hamiltonian vector fields associated with the momenta $I, J$.
Hamiltonian vector fields are  defined in terms of Poisson brackets in the standard way
\be
X_f \equiv \{\cdot \;, f\}  \nonumber
\ee
so that, by indicating with  $X_i= \{\cdot \;, I_i\}, \tilde X^i= \{\cdot \;, \tilde I^i\}$ the Hamiltonian vector field associated with $I_i, \tilde I^i$ respectively, one has, after using the Jacobi identity,  the following Lie algebra:
\beqa
[X_i, X_j] &=& \{\{\cdot \;, I_i\}, I_j\}-\{\{\cdot \;, I_j\}, I_i\}=\{\cdot \; ,\{I_i,I_j\}={ \epsilon_{ij}}^k\{\cdot \; ,I_k\}= {\epsilon_{ij}}^kX_k \nonumber \\
{[}{\tilde X}^i, {\tilde X}^j{] }&=& \{\{\cdot \;, \tI^i\}, \tI^j\}-\{\{\cdot \;, \tI^j\}, \tI^i\}=\{\cdot \; ,\{\tI^i,\tI^j\}\}={ f^{ij}}_k\{\cdot \; ,\tI^k\}= {f^{ij}}_k\tX^k  \nonumber\\
{[}X_i,{\tilde X}^j{] }&=& \{\{\cdot \;, I_i\}, \tI^j\}-\{\{\cdot \;, \tI^j\}, I_i\}=\{\cdot \; ,\{I_i,\tI^j\}\}=  - {f_i}^{ jk}\{\cdot \;,I_k\} - \{\cdot \;, \tI^k\} {\epsilon _{ki}}^{ j} \nonumber\\
&=& - {f_i}^{ jk} X_k -{\tilde X}^k  {\epsilon _{ki}}^{ j}  \nonumber
\eeqa
namely
\be
[X+\tX, Y+\tY]= [X,Y]+ {\sf L}_X \tY-{\sf L}_Y \tX + [\tX, \tY]  \nonumber
\ee
which  shows that  C-brackets can be obtained as  derived brackets, in analogy with  the ideas of  ref.s \cite{deser1, deser2}, with the remarkable difference that, in this case, they are  derived from the canonical Poisson brackets of the dynamics.
 
 \subsection{Poisson-Lie simmetries}\label{PLsym}
 Let us  explicitly address the nature  of symmetries  of the dual models introduced in the previous sections. In particular we want to discuss to what extent the models possess Poisson-Lie symmetries. We closely follow  \cite{marmo:articolo1}  for this subsection.   Poisson-Lie symmetries are Lie group transformations implemented on the carrier space of the dynamics  via group multiplication, which, in general,  are not canonical transformations as they need not preserve
the symplectic structure. However, if the Poisson structure is of the form \eqn{gammagamma2} with  carrier space  $D$ itself,  or \eqn{pbm1}, \eqn{pbm2} if we are looking at  $G$, $G^*$ respectively, Poisson brackets   can be made invariant if  the parameters of the group of transformations are imposed  to have
 nonzero Poisson brackets with themselves. Group multiplication is then said to
correspond to a Poisson map.  We have for example,  for the
right transformations of $G$ on
$D$,
\be\label{rightGact}
\gamma\rightarrow \gamma h \;,\; h \in G\;,\; \gamma \in D
\ee
and the left action of $G^*$ on $D$,
\be\label{leftG*act}
\gamma\rightarrow \tilde h \gamma  \;,\; h^* \in G^*\; \; \gamma \in D
.
\ee
In terms of the coordinates $(\tilde g,g)$ this implies
\be
g\rightarrow gh\;, \quad \tilde g\rightarrow \tilde g \;,
\ee
for the former and
\be
g\rightarrow g\;, \quad \tilde g\rightarrow \tilde h \tilde g  \;,
\ee
for the latter.  By themselves these
transformations  do not preserve the Poisson brackets
\eqn{pbm1}-\eqn{pbm2}.  But they can be made to be invariant
 if we require that the parameters of the tranformation, $h$, have the following Poisson brackets
\be\label{hh}
\{h_1,h_2\}=[\;r^*\;,\;h_1 h_2\;] \;,
\ee
and zero Poisson brackets with $g$ and $\tilde g$.
Then the $SU(2)$ right multiplication is a Poisson map and \eqn{rightGact} corresponds
to a Poisson-Lie group transformation.
For \eqn{leftG*act} to be a Poisson-Lie group transformation, $\tilde h$
must have the following Poisson bracket with itself
\be\label{hsthst}
\{\tilde h_1,\tilde h_2\}=-[\;r\;,\;\tilde h_1 \tilde h_2\;] \;,
\ee
and zero Poisson brackets with $g$ and $\tilde g$.
Since the right-hand-sides of \eqn{hh} and \eqn{hsthst} vanish
in the limit $\lambda \rightarrow 0 $,
the transformations \eqn{rightGact} and \eqn{leftG*act} become canonical in the limit.
\\
Moreover,  Poisson brackets \eqn{pbm1}-\eqn{pbm2}  are invariant under the
simultaneous action of both $G$ and $G^*$ via \eqn{rightGact} and \eqn{leftG*act}, if 
 we assume that{}\be
\{\tilde h_1,h_2\}=0\;.
\ee
By comparing with eq. \eqn{finalpoi} we
conclude that the algebra of the observables
$g$ and $\tilde g$ is different from the algebra of the symmetries
parametrized by $h$ and $\tilde h$.
Therefore, dynamics on the group manifold of $SL(2,\C)$ and on the two partner groups $SU(2)$ and $SB(2,\C)$ possesses  Poisson-Lie group symmetries, when endowed with the above mentioned brackets.  

Let us go back to the symplectic structures of  the IRR and the dual model, respectively given by Eqs. \eqn{xx} and \eqn{xxdu}. The former is obtained  from \eqn{pbm2} while the latter is obtained from  \eqn{pbm1}, for small (but non-zero) value of the parameters $\lambda$ and $\mu$, as we have shown in \ref{canform} (see Eqs. \eqn{calc1}, \eqn{calcdu}). We can therefore conclude that the momentum variables of each model inherit their Poisson brackets from the Poisson-Lie structure  of the dual group, which  in turn exhibits  Poisson-Lie symmetry in the sense elucidated above.

\section{Conclusions and Outlook}\label{concl}
Starting from an existing description of the dynamics of the Isotropic Rigid Rotator on Heisenberg doubles \cite{marmo:articolo1}, we have introduced a new dynamical model which is dual to the standard IRR. To this, we have used the notion of Poisson-Lie groups and Drinfel'd double for understanding the duality between the carrier spaces of the two models. Specifically, we have used the Drinfel'd double of the group $SU(2)$ as the target configuration space for the dynamics of a generalized model, with doubled degrees of freedom. This model exhibits non-Abelian duality and is an ideal arena to analyze in a simple context the meaning to physics of generalized and double geometry structures. 
Moreover, we have shown that, from the generalized action, the usual description with half the degrees of freedom, can be recovered by gauging one of its symmetries. 

The simple model of the IRR is especially interesting as a toy model for field theories with non-trivial target spaces such as Principal Chiral Models. In their original formulation \cite{gursey} these are nonlinear sigma models with the principal homogeneous space of the Lie group $SU(N)$ as its target manifold, where $N$ is the number of quark flavors.  Therefore, the dynamical fields of the model, so called currents take value in the cotangent bundle of the Lie group, while the canonical formalism is described by a Poisson algebra which takes the form of a semi-direct sum. The analogy with the IRR is thus very strict:  the analysis we have performed can be readily generalized, starting from an alternative description of Principal Chiral Models given in ref.s \cite{rajeev:bosonization}, \cite{vitale1, vitale2, delduc} (also see \cite{reid} where Principal Chiral Models are analyzed in the DFT context). 

A Principal Chiral Model is a field theory with target space given by a Lie group $G$ and base space given by the two-dimensional space $\mathbb{R}^2$ endowed with the metric $h_{\alpha \beta}=\mathrm{diag}(-1,1)$. \\
It describes the dynamics of two dimensional fields $g: \mathbb{R}^{1,1}=(\mathbb{R}^2,h) \rightarrow G.$  The action may be written in terms of Lie algebra valued  left invariant one forms 
\be
g^{-1}\mathrm{d}g=g^{-1}\partial_{t}g \mathrm{d}t+g^{-1}\partial_{\sigma}g \mathrm{d}\sigma
\ee
so to have 
\be
S=\frac{1}{2}\int_{\mathbb{R}^2}Tr(g^{-1}\mathrm{d}g \wedge * g^{-1}\mathrm{d}g),\label{chimoac}
\ee
where trace is understood as the scalar product on the Lie algebra $\mathfrak{g}.$  The Hodge operator exchanges the time and space derivatives
\be 
* (g^{-1}\mathrm{d}g)= *( \dot Q^i  dt+{Q^i}' d\sigma ) e_i= ( \dot Q^i  d\sigma-{Q^i}'dt ) e_i
\ee
with $ \dot Q^i =\Tr g^{-1}\partial_{t}g  e_i$, $ { Q^i}' =\Tr g^{-1}\partial_{\sigma}g  e_i$. 
The action \eqn{chimoac} is  the two-dimensional analogue of the IRR action. Notice that in this case the Hodge operator maps one-forms into one-forms while exchanging time and space derivatives. When passing to the Hamiltonian formalism the momenta $I_i= \dot Q^j \delta_{ji}$ and the space derivatives $J^i:= {Q^i}'$ close a Poisson algebra, which, upon an equivalent reformulation of the model \cite{rajeev:bosonization}, \cite{vitale1, vitale2},  results to be  isomorphic to the Kac-Moody algebra $\widehat{\mathfrak{sl}(2,\C)}$. It is therefore natural to conceive a dual model with the same underlying $\widehat{\mathfrak{sl}(2,\C)}$ structure but with the role of $I_i, J^i$ exchanged. The action of the dual model is the natural two-dimensional analogue of \eqn{dualag}, with $\tilde g=\tilde g(\sigma,t)$.  Moreover a parent action encoding both models can be introduced, which is in turn the analogue of \eqn{newac}, with $\gamma=\gamma(\sigma, t)$. The symmetries of the two models under duality transformations are addressed as well. Because of the presence of time and space derivatives that are exchanged by the Hodge operator, the structure is richer than the one exhibited by the particle dynamical systems considered here. We are completing the analysis and  the results will be detailed in a forthcoming paper \cite{MPV2}. 

\noindent{\bf Acknowledgements} 
 P. V.  acknowledges  support by COST (European Cooperation in Science  and  Technology)  in  the  framework  of  COST  Action  MP1405  QSPACE. The authors are indebted with  F. Ciaglia and G. Marmo for  the many invaluable discussions and suggestions all over the preparation of the manuscript. F. P. would like to thank Jeong-Hyuck Park for helpful discussions and Sogang University for their kind hospitality in an early stage of this work. V. Marotta is indebted with R. Szabo for enlightening discussions and useful indications about the existing literature.

\end{document}